\documentclass[twocolumn,twocolappendix]{aastex631}

\usepackage{amsmath}
\usepackage[english]{babel}

\shorttitle{Absorption Systems Toward a Young Weak-line Quasar}
\shortauthors{Andika et al.}

\graphicspath{{./}{figures/}}

\newcommand{\Lya}{Ly$\alpha$}
\newcommand{\PSOJ}{PSO\,J083+11}

\newcommand{\cii}{\ion{C}{2}}
\newcommand{\mgii}{\ion{Mg}{2}}
\newcommand{\mgi}{\ion{Mg}{1}}
\newcommand{\oi}{\ion{O}{1}}
\newcommand{\siii}{\ion{Si}{2}}
\newcommand{\alii}{\ion{Al}{2}}
\newcommand{\feii}{\ion{Fe}{2}}
\newcommand{\civ}{\ion{C}{4}}
\newcommand{\hi}{\ion{H}{1}}

\begin{document}

\title{Staring at the Shadows of Archaic Galaxies:\\
        Damped Ly$\boldsymbol{\alpha}$ and Metal Absorbers Toward a Young $\boldsymbol{z\sim6}$ Weak-line Quasar}

\correspondingauthor{Irham Taufik Andika}
\email{andika@mpia.de}

\author[0000-0001-6102-9526]{Irham Taufik Andika}
\affiliation{Max-Planck-Institut f\"{u}r Astronomie, K\"{o}nigstuhl 17, D-69117 Heidelberg, Germany}
\affiliation{Sterrenkundig Observatorium, Universiteit Gent, Krijgslaan 281 S9, 9000 Gent, Belgium}
\affiliation{International Max Planck Research School for Astronomy \& Cosmic Physics at the University of Heidelberg, Germany}

\author[0000-0003-3804-2137]{Knud Jahnke}
\affiliation{Max-Planck-Institut f\"{u}r Astronomie, K\"{o}nigstuhl 17, D-69117 Heidelberg, Germany}

\author[0000-0002-2931-7824]{Eduardo Ba\~{n}ados}
\affiliation{Max-Planck-Institut f\"{u}r Astronomie, K\"{o}nigstuhl 17, D-69117 Heidelberg, Germany}

\author[0000-0001-8582-7012]{Sarah E. I. Bosman}
\affiliation{Max-Planck-Institut f\"{u}r Astronomie, K\"{o}nigstuhl 17, D-69117 Heidelberg, Germany}

\author[0000-0003-0821-3644]{Frederick B. Davies}
\affiliation{Max-Planck-Institut f\"{u}r Astronomie, K\"{o}nigstuhl 17, D-69117 Heidelberg, Germany}

\author[0000-0003-2895-6218]{Anna-Christina Eilers}
\altaffiliation{NASA Hubble Fellow}
\affiliation{MIT Kavli Institute for Astrophysics and Space Research, 77 Massachusetts Ave., Cambridge, MA 02139, USA}

\author[0000-0002-6822-2254]{Emanuele Paolo Farina}
\affiliation{Gemini Observatory, NSF's NOIRLab, 670 N A'ohoku Place, Hilo, HI 96720, USA}

\author[0000-0003-2984-6803]{Masafusa Onoue}
\affiliation{Max-Planck-Institut f\"{u}r Astronomie, K\"{o}nigstuhl 17, D-69117 Heidelberg, Germany}
\affiliation{Kavli Institute for Astronomy and Astrophysics, Peking University, Beijing 100871, China}
\affiliation{Kavli Institute for the Physics and Mathematics of the Universe (WPI), University of Tokyo, Kashiwa, Chiba 277-8583, Japan}

\author[0000-0002-5027-0135 ]{Arjen van der Wel}
\affiliation{Sterrenkundig Observatorium, Universiteit Gent, Krijgslaan 281 S9, B-9000 Gent, Belgium}

\begin{abstract}

We characterize the \Lya\ halo and absorption systems toward \PSOJ, a unique $z=6.3401$ weak-line quasar, using Gemini Near-infrared Spectrograph, Magellan Folded-port Infrared Echellette, and Very Large Telescope Multi Unit Spectroscopic Explorer (MUSE) data.
Strong absorptions by hydrogen and several metal lines (e.g., \cii, \mgii, and \oi) are discovered in the spectrum, which indicates the presence of: (i) a proximate subdamped \Lya\ (sub-DLA) system at $z=6.314$ and (ii) a \mgii\ absorber at $z=2.2305$.
To describe the observed damping wing signal, we model the \Lya\ absorption with a combination of a sub-DLA with the neutral hydrogen column density of $\log N_\mathrm{HI} = 20.03 \pm 0.30~\mathrm{cm^{-2}}$ and absorption from the intergalactic medium with a neutral fraction of around 10\%.
The sub-DLA toward \PSOJ\ has an abundance ratio of [C/O]~$=-0.04 \pm 0.33$ and metallicity of [O/H]~$=-2.19 \pm 0.44$, similar to those of low-redshift metal-poor DLAs.
These measurements suggest that the sub-DLA might truncate \PSOJ's proximity zone size and complicate the quasar lifetime measurement.
However, this quasar shows no sign of a \Lya\ halo in the MUSE data cube, where the estimated $1\sigma$ limit of surface brightness is $2.76 \times 10^{-18}~\mathrm{erg~s^{-1}~cm^{-2}~arcsec^{-2}}$ at aperture size of 1\arcsec, or equivalent to a \Lya\ luminosity of $\leq 43.46$~erg~s$^{-1}$.
This nondetection, while being only weak independent evidence on its own, is at least consistent with a young quasar scenario, as expected for a quasar with a short accretion timescale.

\end{abstract}


\section{Introduction} 
\label{sec:intro}

Quasars are fueled by the matter accretion onto supermassive black holes (SMBHs).
At high redshifts, they can have SMBH masses of $\gtrsim10^9~M_\odot$ as early as $<1$~Gyr after the Big Bang \citep[][]{2018Natur.553..473B,2020ApJ...897L..14Y,2021ApJ...907L...1W}.
Hence, these quasars are excellent observational probes for understanding the buildup of the first SMBHs and the host galaxies, the early structure formation, and the drivers of cosmic reionization \citep[][]{2020A&A...642L...1M,2021arXiv210803699B,2022MNRAS.509.1885P}.
To date at $z>6$, there are around 300 quasars discovered through various wide-field and deep-sky surveys \citep[e.g.,][]{2006AJ....132..117F,2010AJ....139..906W,2011Natur.474..616M,2015ApJ...801L..11V,2016ApJS..227...11B,2017ApJ...849...91M,2018ApJS..237....5M,2019MNRAS.487.1874R,2019AJ....157..236Y,2019ApJ...884...30W,2019MNRAS.484.5142P,2022ApJS..259...18M}.
Interestingly, tens of them were identified as so-called ``young quasars'' with estimated lifetimes of only $t_\mathrm{Q} \sim 10^4$--$10^5$ yr \citep{2021ApJ...917...38E}.
These young quasars create significant challenges as these luminous quasar lifetimes are many times shorter than the typical time required for early SMBH growth \citep[$t_\mathrm{Q} \sim 10^6$--$10^9$ yr; e.g.,][]{2021MNRAS.505..649K,2021MNRAS.505.5084W}.
The growth of these sources would require either: (i) direct collapse of massive seed black holes \citep[$\gtrsim10^4~M_\odot$; e.g.,][]{2006MNRAS.370..289B,2014MNRAS.443.2410F,2016MNRAS.463..529H,2017MNRAS.471.4878S,2019MNRAS.486.2336D}, (ii) nearly continuous accretion at the (super-)Eddington limit onto lower-mass seeds \citep[$\lesssim 10^{2}$--10$^{3}~M_\odot$; e.g.,][]{2005ApJ...628..368O,2009ApJ...696.1798T,2016MNRAS.459.3738I}, or (iii) the existence of radiatively inefficient accretions \citep[e.g.,][]{2017ApJ...836L...1T,2019ApJ...884L..19D}.
Hence, both expanding the number of known quasars at this early epoch, the search specifically for young quasars, and detailed analyses of their properties are essential for a better understanding of early growth modes and the physics involved.

In our previous work we found a new weak-emission-line quasar at $z = 6.3401$, PSO\,J083.8371+11.8482 \citep[hereafter \PSOJ;][]{2020ApJ...903...34A}.
Inferred from the small size of the proximity zone, we argued that this source clearly belongs to the young quasar population with a lifetime of only $t_\mathrm{Q}\lesssim10^4$ yr for its latest luminous quasar phase. 
However, we found tentative evidence that the presence of a damped \Lya\ system \citep[DLA;][]{2005ARA&A..43..861W} intervening our line of sight to the quasar might spectrally truncate \PSOJ's proximity zone, adding a complication to using the established quasar age measurement approach \citep{2017ApJ...840...24E}. One aim of this paper is to rectify this situation and analyze the potential impact in depth -- as well as to add independent evidence for a young quasar age.

On the other hand, high-$z$ DLAs themselves are of prime interest. 
They are excellent laboratories to examine the neutral-gas reservoirs that give rise to galaxies at cosmic dawn.
These systems have the potential to be good tracers of the metal enrichment history by the first stars and their contribution to the universe's reionization \citep[e.g.,][]{2014ApJ...787...64K,2017MNRAS.472.3532M}.
Moreover, metal-poor DLAs at high redshifts are considered to be the progenitor of modern-day dwarf galaxies, which occupy the galaxy luminosity function at the faint end \citep{2015ApJ...800...12C}.
Finding and characterizing $z\gtrsim6$ DLAs becomes extremely difficult due to increasing \Lya\ forest opacity \citep{2018ApJ...864...53E,2018MNRAS.479.1055B} of the intergalactic medium (IGM).
Although metal absorbers are frequently detected at $z\gtrsim6$, their hydrogen content is often unknown so that the absolute metal abundances cannot be determined \citep[e.g.,][]{2019ApJ...882...77C}.
The currently best approach is to study ``proximate'' DLAs that reside close, in redshift, to a background quasar -- i.e., within $\sim$5000~km~s$^{-1}$ -- so that their \Lya\ damping wing absorptions extend into the forest-free quasar continuum, allowing us to estimate their neutral hydrogen column density ($N_\mathrm{HI}$).
To date, there are only four proximate DLAs found at $z\gtrsim6$: 
SDSS\,J2310+1855 \citep[$z_\mathrm{DLA} = 5.939$;][]{2018ApJ...863L..29D}, 
PSO\,J056--16 \citep[$z_\mathrm{DLA} = 5.967$;][]{2020MNRAS.494.2937D}, 
PSO\,J183+05 \citep[$z_\mathrm{DLA} = 6.404$;][]{2019ApJ...885...59B}, 
and P007+04 \citep{2022Farina}.

In this work, we present a study of the environment and absorption systems toward \PSOJ.
Using new integral-field unit (IFU) spectroscopic data from the Very Large Telescope (VLT) Multi Unit Spectroscopic Explorer (MUSE) together with deep near-infrared (NIR) spectroscopy presented in \cite{2020ApJ...903...34A}: (i) we report on a newly discovered $z=6.314$ proximate absorber in the sightline of this quasar and constrain its effects on the quasar's lifetime estimation; and (ii) we investigate the existence of a \Lya\ extended emission around the quasar itself that would add knowledge to the current quasar lifetime.

The structure of this paper is as follows.
We start in Section~\ref{sec:spec_obs} by describing data acquisition and reduction.
Section~\ref{sec:damping_wing} describes our metal absorption lines and \Lya\ damping wing measurements.
After that, Section~\ref{sec:halo_find} presents our point-spread function modeling and methods to subtract out quasar continuum light to set limits on a \Lya\ halo around the quasar. 
Furthermore, we discuss the results in Section~\ref{sec:discussion}, including the elemental abundances of this proximate absorber and whether or not the absorber impacts the interpretation of the quasar's observed proximity zone.
We close with a summary of the results and our conclusions in Section~\ref{sec:summary}.

For all calculations, we use the flat $\Lambda$CDM cosmology, with $\Omega_\Lambda=0.7$, $\Omega_\mathrm{m}=0.3$, and $H_0 = 70~\rm km~s^{-1}~Mpc^{-1}$.
As a result, at $z = 6.3401$, the age of the universe is 0.852~Gyr, and the angular scale of $\theta = 1\arcsec$ corresponds to proper transverse separation of 5.5~kpc.

\section{Observations and Data Reduction}
\label{sec:spec_obs}

As stated before, our goals are twofold. 
First, we want to make a deeper analysis of the quasar's \Lya\ damping wing and its potential impact, and then also the physical properties of a potential intervening proximate absorber creating strong metal absorption lines, based on NIR spectroscopy and modeling.
The second goal is testing the prediction of the ``young quasar'' picture, which is the limited presence of extended \Lya\ emission near the central quasar. For this purpose, we obtained VLT/MUSE integral-field unit (IFU) spectroscopy to analyze the environment of \PSOJ\ that is spatially and spectrally resolved.
The details of discovery and characterization of \PSOJ\ as well as Hubble Space Telescope and Atacama Large Millimeter/submillimeter Array (ALMA) data, and the results were presented in \citet{2020ApJ...903...34A}.
Here we recap the main properties of the initial and follow-up NIR spectroscopy, followed by a description of the VLT/MUSE IFU data.

\subsection{Initial Observing Run with the Magellan Folded-port Infrared Echellette} \label{sec:fire_spec}

The first NIR spectroscopy of \PSOJ\ was obtained in January and February 2019 utilizing the 6.5m Magellan Folded-port InfraRed Echellette (FIRE) instrument \citep[PI: R.~Simcoe;][]{2013PASP..125..270S}.
The instrument was configured to the high-resolution echellette mode using the 0\farcs6 slit to observe the target quasar for 5~hr.
The resulting spectral data have a resolution in velocity space of $\sim50$~km~s$^{-1}$, equivalent to a spectral resolution of $R \sim 6000$ within the wavelengths of 0.82--2.51~$\mu$m.
Unfortunately, this observing run was conducted in suboptimal weather conditions, which results in a degraded signal-to-noise ratio (S/N).
Nonetheless, the data are well suited for metal line diagnostic due to their spectral resolution, as shown below.

\subsection{Gemini Near-infrared Spectroscopy} \label{sec:gnirs_spec}

For proper characterization of the \Lya\ damping wing, we performed a second spectroscopic campaign to both create a high S/N -- but lower resolution -- spectrum in that region and at the same time calibrate out instrument-specific effects at the long-wavelength end.
This second run was performed on 20--22 March 2019 with 8060~s total time of integration on target using the 8.1 m Gemini-N Near-infrared Spectrograph (GNIRS; GN-2019A-FT-204, PI: M.~Onoue). We chose the cross-dispersed mode to encompass a wavelength range of $\lambda_{\rm obs}\sim$ 0.9--2.5~$\mu$m in the observed frame, using a 31.7~l/mm grating and a `short' camera with 0\farcs15 per pixel resolution.
Using an aperture size of 0\farcs675 slit resulted in a spectral resolution of $R\sim750$.
Single-frame exposure was 155~s long, and between exposures, a canonical ABBA pattern was used to reduce the noise from skylines.
The observations were taken at an air mass range of $\sim1.1$--1.7.

The details of the spectroscopic data reduction can be found in \citet{2020ApJ...903...34A}. 
In summary, \texttt{PypeIt}\footnote{\url{https://pypeit.readthedocs.io/en/latest/}} \citep{2020JOSS....5.2308P} was used from cleaning the raw 2D spectrum -- image differencing, flat-fielding, cosmic-ray removal, etc. -- to producing the wavelength- and flux-calibrated 1D spectrum.
After that, contamination from telluric absorptions was corrected using \texttt{Molecfit}\footnote{\url{https://www.eso.org/sci/software/pipelines/skytools/molecfit}} \citep{2015A&A...576A..78K,2015A&A...576A..77S}.
Finally, we utilized the dust map of \citet{2019ApJ...887...93G} and extinction relation from \cite{2016ApJ...826..104G} to correct reddening due to Galactic extinction.

\subsection{IFU Spectroscopy with VLT/MUSE} \label{sec:muse_spec}

Specifically for the proximate absorber and environment analysis, we added red optical IFU spectroscopy for \PSOJ\ with MUSE at the 8.2m ESO VLT (0104.B-0665(A), PI: Andika), using the instrument's wide-field mode.
The resulting data cube has a spectral resolution of $R \sim 4000$ and covers the wavelength range of 0.47--0.93~$\mu$m.
The quasar was observed for 3~hr, divided into five exposures of 2116~s, with shifts of $<5$\arcsec\ and rotations of 90$^\circ$ between exposures.
Around the observed wavelength of \Lya\ the point-spread functions of bright stars in the field have a median width of 0\farcs5.
For data reduction, we used the \texttt{MUSE DATA REDUCTION SOFTWARE} version 2.6 \citep{2012SPIE.8451E..0BW, 2014ASPC..485..451W}, complemented with a pipeline developed by \citet{2019ApJ...887..196F}.
The individual exposure was rescaled in flux before combining to avoid possible fluctuation of the photometry due to different weather conditions.
Then the variance data cube was rescaled to match the observed background's variance.
Next, we improve the astrometric calibration by anchoring the sources to the Panoramic Survey Telescope and Rapid Response System 1 catalog \citep{2016arXiv161205560C} and correct the Galactic reddening. 
After that, the contamination from night skylines was corrected utilizing the \texttt{Zurich Atmospheric Purge} software \citep{2016MNRAS.458.3210S}.
Finally, the spectrum of the quasar is extracted from the data cube using a circular aperture with a radius of 0\farcs75.

\section{What shapes the Lyman-alpha damping wing?} \label{sec:damping_wing}

The spectral data from MUSE, FIRE, and GNIRS allow us to investigate the presence of an absorber -- e.g.,\ (sub-)DLA or Lyman Limit System\footnote{
Following \cite{2019ApJ...882...77C}, we employ the \hi-based definition to classify the absorber as a Lyman Limit System ($\log N_\mathrm{HI} \geq 17.2$~cm$^{-2}$), sub-DLA ($19.0 < \log N_\mathrm{HI} < 20.3$~cm$^{-2}$), or DLA ($\log N_\mathrm{HI} \geq 20.3$~cm$^{-2}$).
} -- close to the quasar that might influence the \Lya\ damping wing.
In fact, we identified line absorption that we associated with the \mgii~$\lambda\lambda2796,2803$ doublet at $\lambda_\mathrm{obs} = 20\,449.9, 20\,501.1$~\AA; \cii~$\lambda1334$ at $\lambda_\mathrm{obs} = 9756.9$~\AA; and a marginal detection of \oi~$\lambda1302$ at $\lambda_\mathrm{obs} = 9522.8$~\AA.
This is a strong indication of the proximate absorber presence located at $z=6.314$.
On the other hand, we could not confidently identify other associated metal line absorption like \siii, \alii, and \feii\ that we would expect for a specifically strong DLA system.
This indicates that the aforementioned proximate system is likely not a particularly strong absorber.
In this section, we will constrain the quasar's column density of neutral hydrogen to see whether the absorption from the proximate absorber is a dominant or significant factor impacting the \Lya\ damping wing detected in the \PSOJ\ spectrum.

As an aside, we also found independent line signatures of a strong \mgii\ absorber at a lower redshift -- i.e., at $z = 2.2305$ -- which we identified from the absorptions of \mgii~$\lambda\lambda2796,2803$ doublet and \mgi~$\lambda2853$ at $\lambda_\mathrm{obs} = 9032.5,\ 9055.1,\ \mathrm{and}\ 9216.6$~\AA, respectively.

\subsection{Metal Absorption Line Analysis} \label{sec:absline_analysis}

\begin{figure*}[htb!]
	\centering
	\epsscale{1.17}
	\plottwo{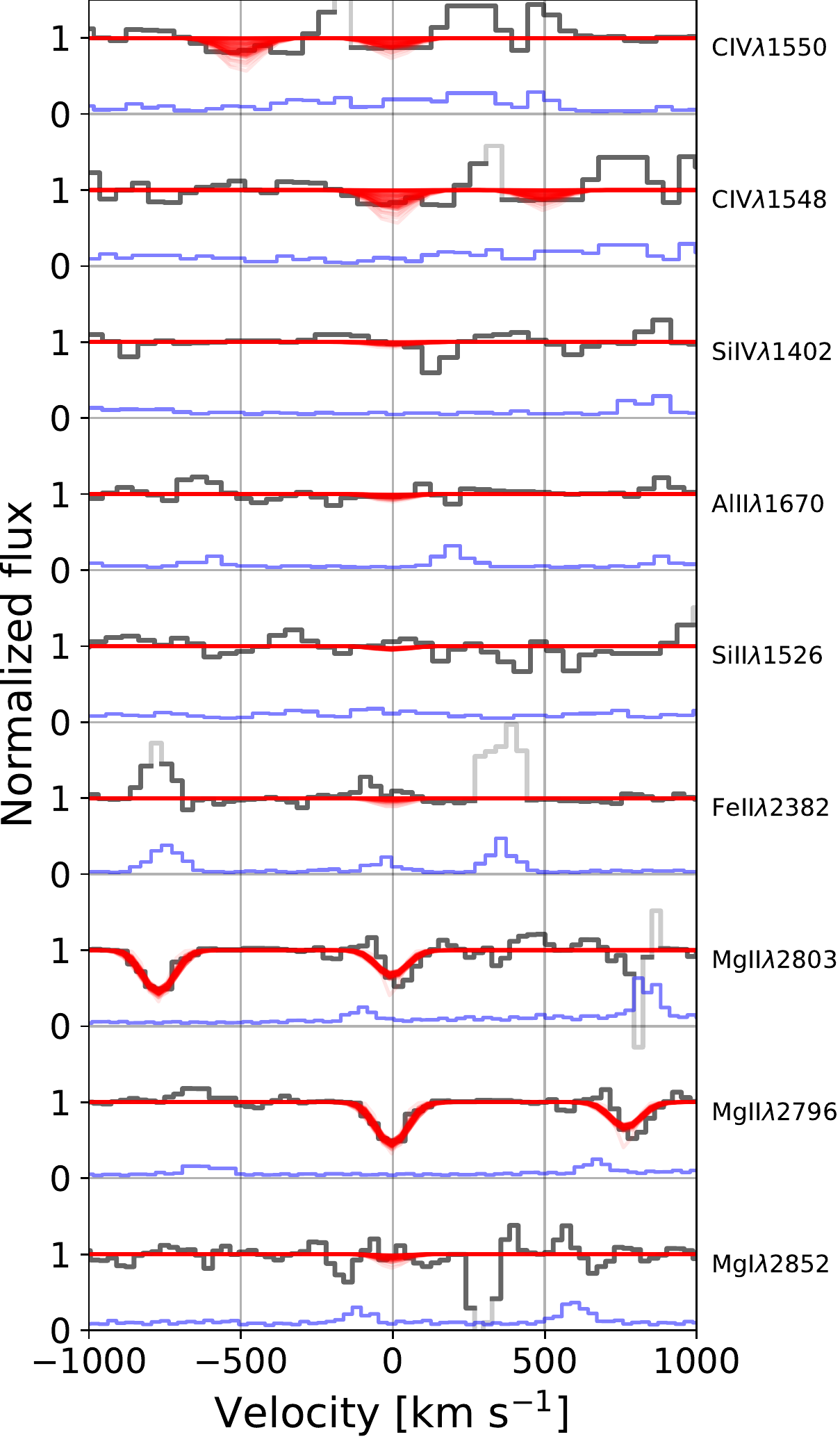}{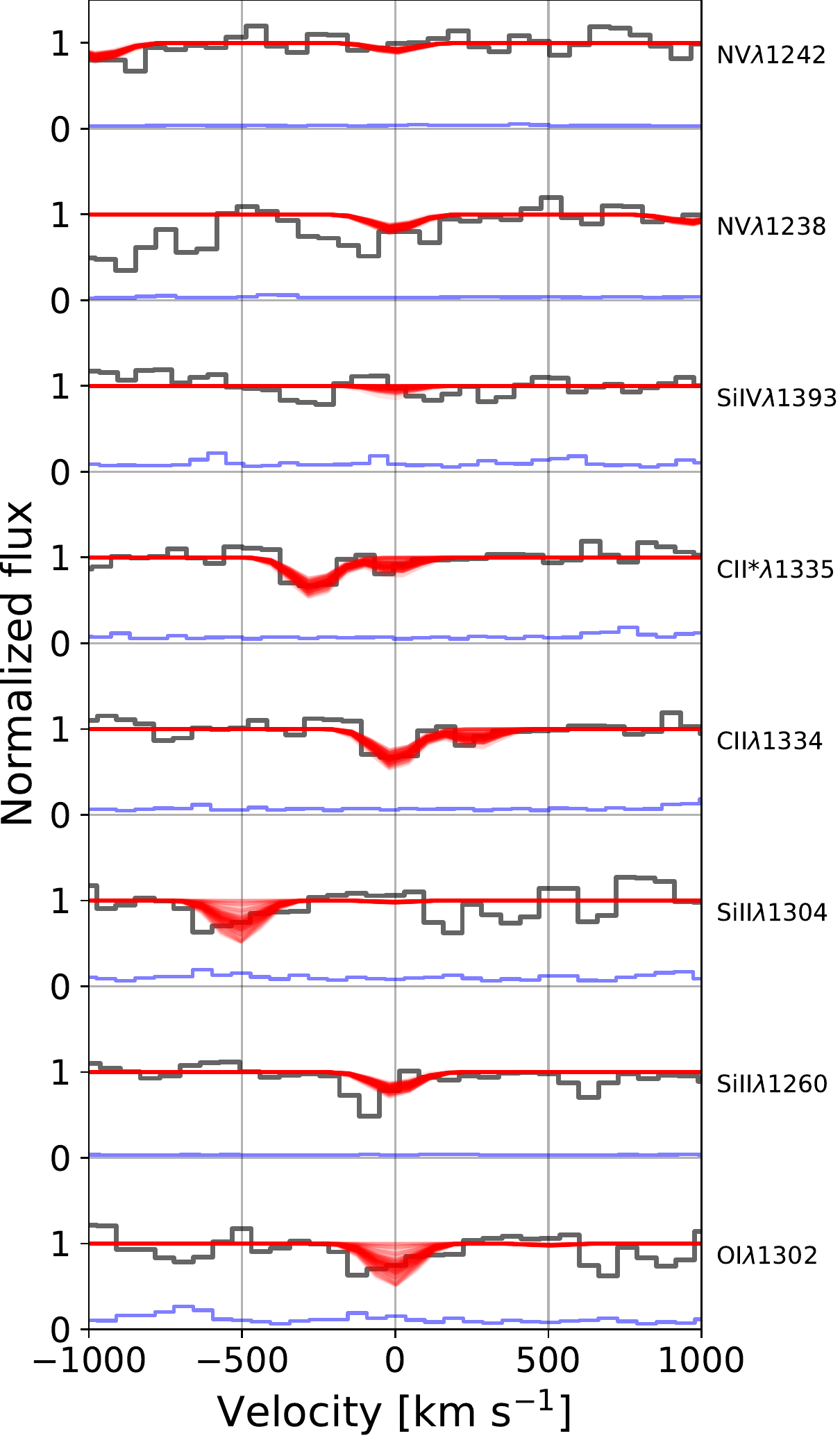}
	\caption{
		Velocity stack of the sub-DLA metal absorption lines at $z=6.314$ observed in the continuum-normalized combined MUSE + FIRE spectrum.
		The data are shown as black lines with fitted models superposed in red.
		Note that the width of the red lines corresponds to the model uncertainty (see Section~\ref{sec:absline_analysis}).
		Blue curves show a $1\sigma$ shot noise spectrum.
		We use gray lines to show the data in regions with high noise due to skylines.
		The absorptions from \mgii~$\lambda\lambda2796,2803$ doublet and  \cii~$\lambda1334$ are clearly seen while there is only a marginal detection of \oi~$\lambda1302$.
        Finally, the other associated metals like \siii, \alii, and \feii, which one would expect specifically for a strong absorber, are not apparent in the spectrum.
	}
	\label{fig:absline}
\end{figure*}

The spectra we utilized for metal absorption line measurements in this section are created based on the combined MUSE and FIRE data.
As mentioned before, the FIRE data were taken during suboptimal weather conditions, which degrades its S/N.
Hence, we first bin and convolve the FIRE spectrum to have a similar spectral resolution as MUSE (i.e., $R\sim4000$); as a consequence, the signal in the data also increases.
Then, we stitch them together by keeping the data from MUSE at the observed wavelength of $\lambda_\mathrm{obs} \leq 9300$~\AA\ while using the FIRE spectrum at $\lambda_\mathrm{obs} > 9300$~\AA.

To calculate the column densities of the metals presented in Figure~\ref{fig:absline}, we needed to continuum normalize the \PSOJ\ spectrum first. We did this under the general assumption that the bulk of high-frequency structure were either absorption lines or was noise. 
With this assumption, the continuum was modeled using \texttt{QSmooth}\footnote{\url{https://github.com/DominikaDu/QSmooth}} \citep{2020MNRAS.493.4256D}: this code first bins the spectrum with a running median of 50 data points to capture a first rough estimate of the main continuum and potential broad emission lines.
It then constructs the spectrum's upper envelope by performing a peak-finding procedure above the aforementioned running median and then interpolates the peaks.
This envelope is almost independent of deep absorption features. It is then subtracted from the spectrum to create a first estimate of a continuum-subtracted spectrum -- although while disregarding the impact of noise.
\texttt{QSmooth} then applies the ``Random Sample Consensus'' regressor algorithm \citep{10.1145/358669.358692} to search for statistical outlier in this intermediate subtracted spectrum. This process decides what is the inherent noise structure in the residual peak-subtracted continuum while rejecting most absorption lines. 
The ``inliers'' data points flagged by the algorithm -- i.e., not being part of emission or absorption features -- are interpolated and smoothed again by calculating a running median. This remaining structure is again removed, hence producing the final smooth flux fit of the spectrum.

Visual inspection is carried out on the continuum-normalized spectrum to check for the detection of potential other metal lines belonging to the very obvious \mgii\ at $z=6.314$.
The column densities were estimated straight from the spectral data by utilizing a \texttt{Python} code developed by \citet[see their Section 3]{2020arXiv201110582S}.
The code employs Markov Chain Monte Carlo (MCMC) samplers to explore the probability distributions of column density, the effect of saturation, and the degeneracies in the fitting parameters.

Before modeling, the user needs to provide a hierarchical absorption \texttt{Model} class.
This class may consist of one or more fitting \texttt{Component} variables, representing absorbers with single or multiple-velocity clouds.
Then, for each \texttt{Component} one needs to specify priors on the redshift ($z$), thermal parameter ($T$) -- which relates to the temperature -- and turbulence parameter ($b$).
Each \texttt{Component} contains several \texttt{Ion} children, where each \texttt{Ion} has a name (e.g., \mgii), a prior on column density ($N$), and an associated dictionary of \texttt{Transitions}.
The \texttt{Transitions} themselves represent absorption lines and are defined based on their rest wavelengths and corresponding atomic data \citep{2003ApJS..149..205M}.

The user then creates a \texttt{Model} class by entering \texttt{Component}, \texttt{Ion}, and \texttt{Transitions} along with suitable wavelength ranges to be modeled.
In general, we fit regions within a velocity space of $\Delta v = \pm 125$~km~s$^{-1}$ around the center of each line, with some modifications for regions with higher noise.
Note that the \texttt{Model} class contains a built-in method to generate Voigt profiles and convolve them with the appropriate instrumental resolution.
This hierarchical workflow naturally fits the absorption component, which contains multiple ions and column densities, with a single value of $z$, $b$, or $T$.
Furthermore, a single value of $N$ can also be used for all transitions of an ion.

In our case, we employ a model using a single velocity component with column densities measured for 11 different ions containing 14 fitting parameters.
This model and its corresponding priors are then supplied to the \texttt{emcee} software \citep{2013PASP..125..306F} to perform MCMC sampling and evaluate the posterior distributions of parameters.
We note that flat priors are used for all input parameters, where $11 \leq \log N \leq 17$, $3 \leq \log T \leq 5$, and $2 \leq b \leq 100$~km~s$^{-1}$.
The code is also able to estimate the upper limit of $N$ because the posterior ranges from the prior lower limit to the maximum $N$ allowed by the observed spectrum.
In the case of undetected ions/lines, the $z$ and $b$ parameters are constrained by other ions in the same velocity component with significant detections.
The reported upper limits for nondetections mean the value below where 95\% of the posterior distribution is found.

Figure~\ref{fig:absline} displays the models fitted to the MUSE + FIRE spectral data.
A sample of 100 models chosen randomly based on the posterior distribution is shown with the red lines -- the width of the red lines hence corresponds to the model uncertainty.
The resulting fitting parameters are listed in Table~\ref{tab:metallicity}, third column.
Here, the reported abundances of C, O, Mg, Al, Si, and Fe are based on derived \cii, \oi, \mgii, \alii, \siii, and \feii\ column densities, respectively.
We report a median and its corresponding 5\% and 95\% confidence levels.
For undetected lines, we write a 95\% upper limit.
The posterior distributions of $T$, $b$, and the metal column densities are reported in Appendix~\ref{sec:appendix_c}.
The model we prefer consists of a proximate absorber at $z=6.314$ with thermal parameter of $\log T = 4.13 \pm 0.80$~K and turbulent parameter of $b = 59 \pm 20$~km~s$^{-1}$.
We note that most of the high-excitation lines such as \civ, \ion{Si}{4}, and \ion{N}{5} are undetected within the sensitivity limits of our data. 
Therefore, the spectra with higher resolution and S/N would be required to give better constraints on those ions.

As a piece of additional information, the previously identified \mgii\ absorber at lower redshift is best modeled with $z=2.2305$, $\log T = 3.99 \pm 0.69$~K, and $b = 56 \pm 16$~km~s$^{-1}$.
This absorber also has column densities of $\log N_\mathrm{MgI} = 12.86 \pm 0.07~\mathrm{cm^{-2}}$ and $\log N_\mathrm{MgII} = 13.75 \pm 0.04~\mathrm{cm^{-2}}$.

\subsection{Modeling the Lyman-alpha Line} \label{sec:damping_fit}

\begin{figure*}[htb!]
	\centering
	\epsscale{1.17}
	\plotone{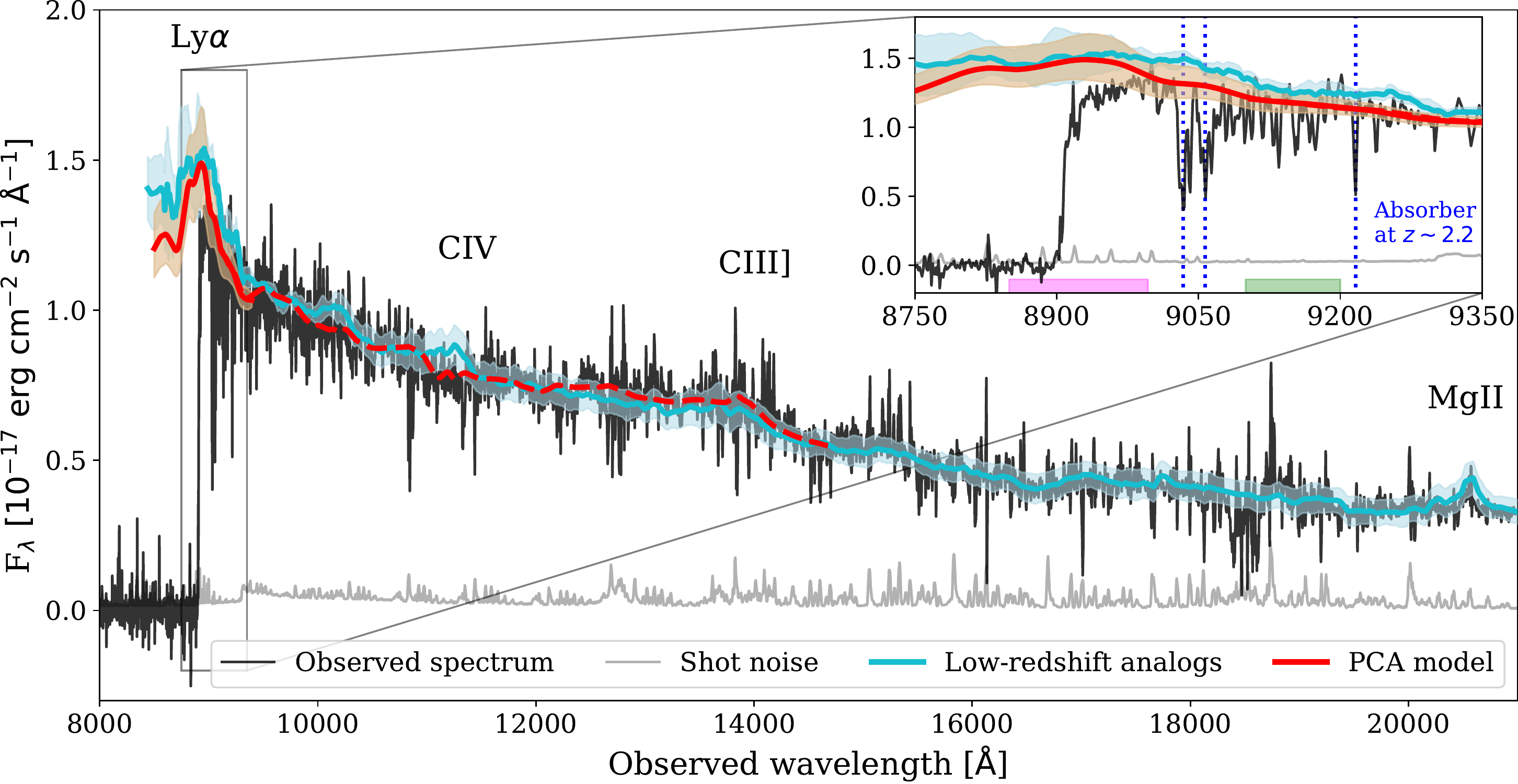}
	\caption{
		The MUSE + GNIRS spectrum of \PSOJ\ (black line) and its associated shot noise (gray line).
		The median composite spectrum of the low-redshift quasar analogs is shown as a cyan line with the $1\sigma$ dispersion around the median as a light blue region.
		The composite spectrum gives a decent fit to the main features observed in the \PSOJ\ spectrum redward of the \Lya\ emission.
		On the other hand, the PCA model to better predict the blue side of the quasar spectrum and its $1\sigma$ dispersion are denoted with the red line and shaded region, respectively.
		The wavelength range and spectrum used in the PCA fit to predict \Lya\ is shown with the red dashed line.
        The inset figure shows the zoom-in to the region around \Lya.
        The \mgii~$\lambda\lambda2796,2803$ doublet and \mgi~$\lambda2853$ absorptions at $z=2.2305$ are marked with blue dotted lines. 
        The wavelength ranges highlighted with magenta ($\lambda_\mathrm{obs}=$~8850--8997~\AA) and green ($\lambda_\mathrm{obs}=$~9100--9200~\AA) colors are the region of interest for creating the \Lya\ pseudo-narrowband and continuum images, respectively.
	}
	\label{fig:spec_model_wide}
\end{figure*}

\begin{figure*}[htb!]
	\centering
	\epsscale{1.17}
	\plottwo{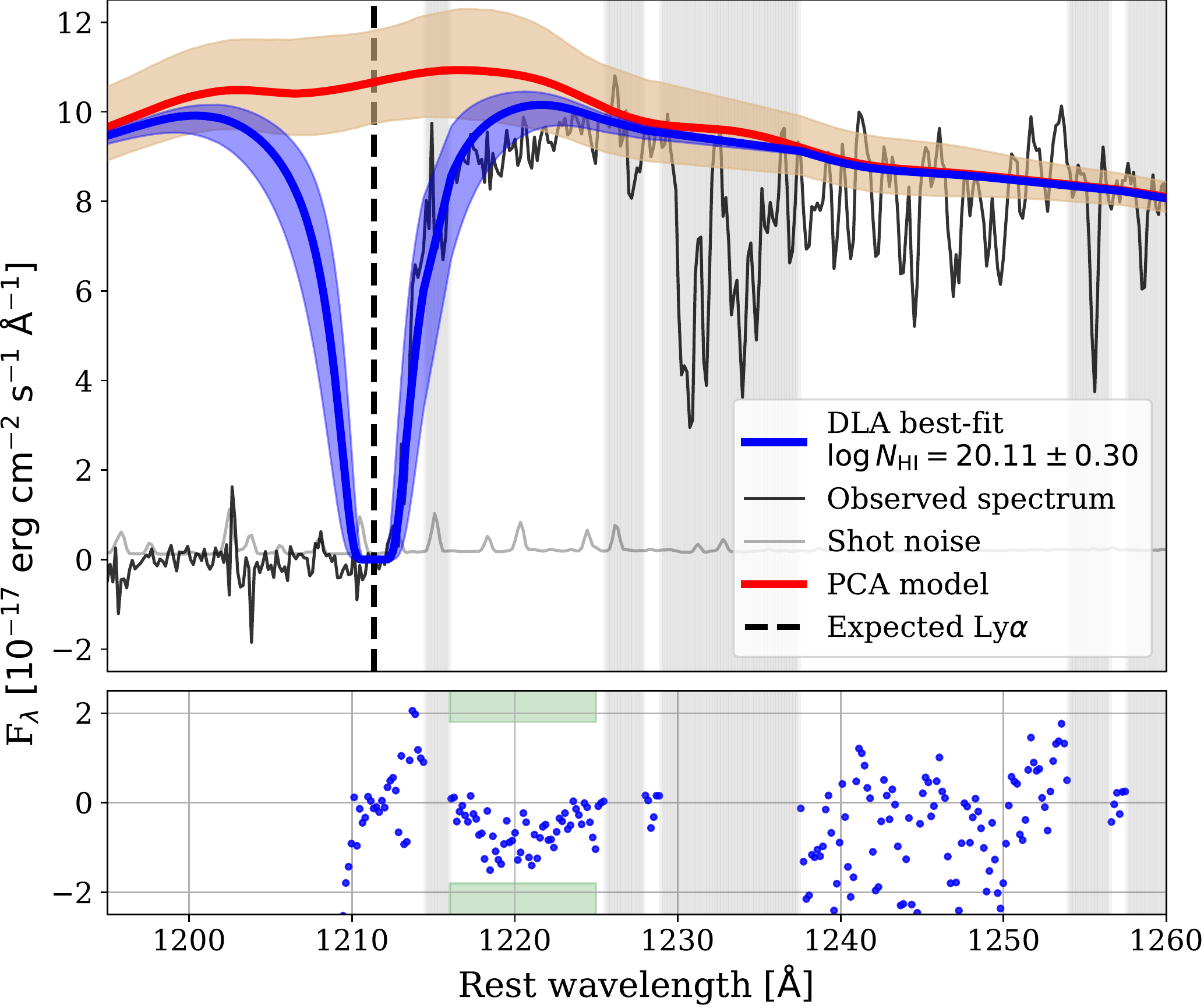}{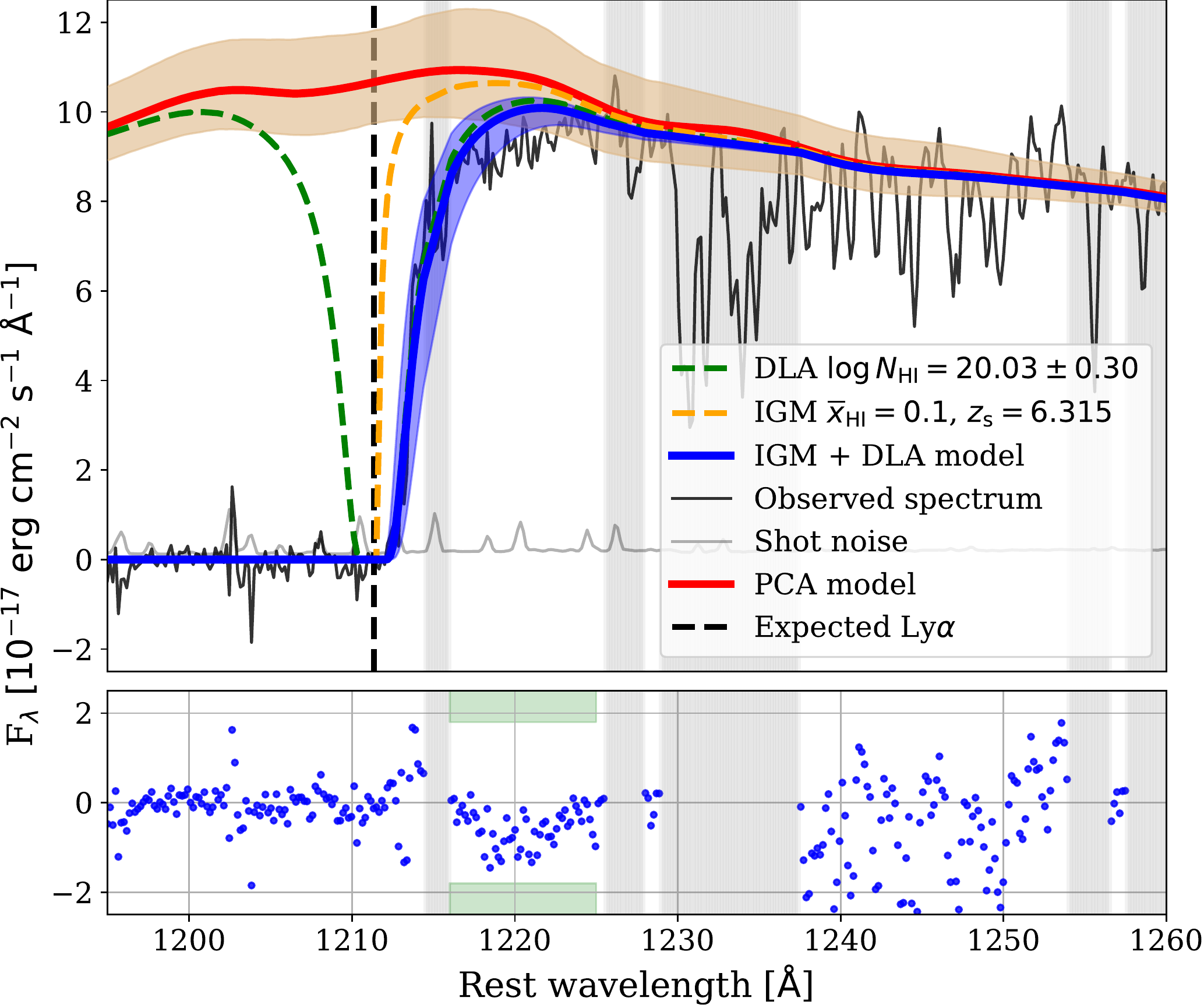}
	\caption{
		Spectrum of \PSOJ\ (black line), shot noise (gray line),  and associated PCA model (red line and shaded region) around the wavelength region where the quasar's \Lya\ emission is expected.
    	The \textit{left panel} shows the absorption from a $z=6.314$ sub-DLA with a hydrogen column density of $\log N_\mathrm{HI} = 20.11 \pm 0.30$ cm$^{-2}$ (blue line and shaded region).
    	Even though it already gives a good fit, we explore an alternative model to test how much the \PSOJ's proximity zone is truncated due to the presence of the aforementioned sub-DLA.
		The \textit{right panel} presents a joint model where we incorporated attenuation due to an IGM with $\overline{x}_\mathrm{HI} = 0.1$ (orange dashed line) and sub-DLA with $\log N_\mathrm{HI} = 20.03 \pm 0.30$ cm$^{-2}$ (green dashed line).
        This joint IGM + DLA model (blue line and shaded region) reconstructs the data at $\lambda = $1216--1225~\AA\ slightly better, although not significant (see wavelengths marked with green). 
        The residual of the fits (blue circles) and masked wavelengths to avoid strong absorption lines or regions with large uncertainty in the sky subtraction (gray shaded area) are also presented in the figure.
		See Appendix~\ref{sec:appendix_a} for the impact of other $\overline{x}_\mathrm{HI}$ values.
	}
	\label{fig:spec_model}
\end{figure*}

How strong is the \Lya\ absorption from the intervening proximate absorber in the line of sight of \PSOJ? To estimate this, we needed to model \PSOJ's intrinsic \Lya\ emission first.
This was done with two independent approaches: (i) stacking the spectra of lower-redshift quasars that have similar emission-line properties as \PSOJ\ to create a template \citep[e.g.,][]{2019ApJ...885...59B} and (ii) constructing a model based on principal component analysis \citep[PCA; e.g.,][]{2018ApJ...864..143D,2021MNRAS.503.2077B}.
For this analysis, we used the MUSE + GNIRS spectrum because although GNIRS has a lower resolution compared to FIRE, it gives better continuum S/N in the spectral data, which is useful for modeling the intrinsic quasar emission.
The combined spectrum is then created by stitching the $\lambda_\mathrm{obs} \leq 9300$~\AA\ MUSE data to the $\lambda_\mathrm{obs} > 9300$~\AA\ GNIRS spectrum.

For the first approach, the construction of the empirical composite spectra based on low-redshift analogs relied on the knowledge that there is no significant quasar spectral evolution across redshifts \citep[e.g.,][]{2019ApJ...873...35S}.
However, the $z\gtrsim6$ quasars often show high-ionization broad lines with more extreme velocity shifts compared to other $z<5$ quasars at the same luminosity \citep{2017ApJ...849...91M,2019MNRAS.487.3305M,2020ApJ...905...51S}.
Nevertheless, a large number of Sloan Digital Sky Survey \citep[SDSS;][]{2018ApJS..235...42A} quasars at lower-$z$ provided a reasonable reference for creating a composite spectrum.
Our method can be summarized as follows:
\begin{enumerate}
	\item All quasars from the SDSS Data Release (DR) 14 quasar catalog \citep{2018A&A...613A..51P} flagged as nonbroad absorption lines (\texttt{BI\_CIV} = 0~km~s$^{-1}$) were retrieved. We limited the selection to the redshifts of $2.0 < z < 2.5$ to include the \Lya, \civ, and \mgii\ lines in the spectra. This yielded 85,535 quasars.
	
	\item The wavelength region around \civ\ was fit with a power law to model the local continuum \citep[for reference see][]{2011ApJS..194...45S}. Then, we estimated the \civ\ equivalent widths (EWs) based on the excess flux above the continuum over the wavelength range where \civ\ is expected.
	In this way we selected quasars with a \civ\ EWs similar with \PSOJ\ (EW~$\lesssim 5.8$ \AA). These criteria left us with 23 quasar ``analogs'' with spectral properties similar to \PSOJ.
	
	\item The spectra of these analogs are smoothed by using \texttt{QSmooth} \citep[see above]{2020MNRAS.493.4256D} to remove specifically noisy regions and strong absorption lines in the spectra.
	We then normalized each of the 23 spectra at 1290~\AA\ before averaging them into our final template.
	
\end{enumerate}
The median composite spectrum and its $1\sigma$ dispersion are presented in Figure~\ref{fig:spec_model_wide}.
The composite spectrum matches most of the traits observed in the \PSOJ\ spectrum and predicts an intrinsically weak \Lya\ line.

For the second approach, we used PCA of lower-redshift quasar spectra to predict the strength and shape of \Lya\ using strong correlations between \Lya\ emission and other rest-frame ultraviolet broad emission lines that are known to be present \citep{1992ApJ...398..476F,2004AJ....128.2603Y,2006ApJS..163..110S}.
This makes PCA approach to also predict the blue part ($\lambda_\mathrm{rest} < 1290$~\AA) of the quasar spectrum -- including \Lya\ -- based only on its red part \citep[$\lambda_\mathrm{rest} \geq 1290$~\AA; e.g.,][]{2006ApJS..163..110S,2011A&A...530A..50P,2018ApJ...864..143D,2018ApJ...864..142D}.
We refer to \cite{2021MNRAS.503.2077B} for the details on the PCA model construction that we used.
In brief, our training set consisted of $\sim4000$ quasars again retrieved from the SDSS DR14 quasar catalog \citep{2018A&A...613A..51P}.
Then, PCA decomposition was performed to capture 70--80\% of the total spectral variance with a linear combination of only 10 and 6 principal components for the red- and blue side spectra, respectively.
After that, we calculated a projection matrix that connects the coefficients from the red side to those on the blue side of the spectrum.
Thus, the blue side spectrum can be predicted based on the projected blue side coefficients and the associated spectral template.
The PCA model for \PSOJ\ is presented in Figure~\ref{fig:spec_model_wide}.
It predicts a weaker \Lya\ and provides a formally better match to the observed data than the constructed lower-$z$ ``analogs'' template, although both models are still consistent with each other within the estimated uncertainties.
Compared to the composite spectrum based on the low-redshift quasar analogs, however, PCA considers not only \civ\ but implicitly also all broad lines characteristics, which makes it in principle superior given the extra information it uses.
Therefore, we utilize the PCA model for the remainder of the analysis to take advantage of its well-quantified uncertainties.

This then allowed us to model the absorption in the \Lya\ region. 
A Voigt model centered at $z = 6.314$ was fitted to the observed damping wing in the \PSOJ\ spectrum -- we chose this proximate absorber redshift based on the location of metal absorption lines (see Section~\ref{sec:absline_analysis}).
Optimization using least squares yielded a best-fit value of $\log N_\mathrm{HI} = 20.11 \pm 0.30$ cm$^{-2}$.
An uncertainty in the range of $N_\mathrm{HI}$ was estimated by overplotting Voigt profiles, modifying the column density input to determine the allowed range by the observed spectrum and the continuum model.
This approach to making a subjective visual estimate of the uncertainty in the absorption profile is currently the standard methodology in the field \citep[e.g.,][]{2019ApJ...885...59B}, because it is difficult to quantify the full set of errors from continuum mismatch, unaccounted-for absorption lines, and shot noise. 
Hence making a statement on the goodness of fit just from $\chi^2$ alone would create an unrealistically small range of likely consistent $N_\mathrm{HI}$ values.
Note that although a single Voigt profile already gives a reasonable fit overall, we want to explore an alternative model to test how much the \PSOJ's proximity zone is truncated due to the presence of the aforementioned proximate absorber.

Another mechanism that is likely involved and that influences the damping wing is the hydrogen absorption in the IGM with a neutral fraction of $\overline{x}_\mathrm{HI}>10\%$ \citep{1998ApJ...501...15M,2019ApJ...885...59B}.
Following the formalism of \citet{1998ApJ...501...15M}, we tried to model the IGM damping wing presuming a constant neutral fraction from quasar's proximity zone at redshift $z = z_\mathrm{s}$ to $z=5.5$, while being entirely ionized around $z\lesssim5.5$.
By definition, the proximity zone is the physical radius at which the transmitted flux drops to 10\%, which for \PSOJ\ is equivalent to $R_\mathrm{p} = 1.17\pm 0.32$~Mpc \citep{2020ApJ...903...34A}.

As discussed by \citet[see their Appendix A.2]{2019ApJ...885...59B}, fitting the data with an IGM + DLA combined model with three free parameters ($\overline{x}_\mathrm{HI}$, $N_\mathrm{HI}$, and $R_\mathrm{p}$) would give a highly degenerated result.
To reduce this dimensionality problem, we stepped through several IGM damping wing appearances using a grid of constant $\overline{x}_\mathrm{HI}$ and $R_\mathrm{p}$.
Then, we fitted the DLA Voigt profile to the already attenuated continuum.
The result is that indeed a joint model of IGM plus DLA always gives a slightly better, although not significant, fit than that using only a DLA, especially at $\lambda=$~1216--1225~\AA\ (see Figure~\ref{fig:spec_model}). 
For the \Lya\ emission model reconstructed with PCA, we found that the allowed DLA profiles that can produce the observed damping wing have $\log N_\mathrm{HI}=$~20.03--19.73~cm$^{-2}$ for an IGM that is 10--50\% neutral.
Consequently, this exercise also gives the proximity zone size that is allowed by the observed spectrum, i.e., $R_\mathrm{p} \gtrsim1.35$~Mpc.
Note that we can only calculate the lower limit of $R_\mathrm{p}$ because if the quasar's proximity zone extends beyond the DLA location (i.e., $z_\mathrm{s} < z_\mathrm{DLA}$), it will be hidden from the observer and poorly constrained due to the blockage by the DLA cloud.
Furthermore, the proximity zone size will no longer be sensitive to quasar lifetime, and hence another independent diagnostic for measuring the age would be required.
See Appendix~\ref{sec:appendix_a} for a display of how different combinations of the neutral fractions and $z_\mathrm{s}$ affect the damping wing model.

For the remainder of the analysis, we choose our preferred model, which is a 10\% neutral IGM and corresponds to the best-fit DLA Voigt profile of $\log N_\mathrm{HI} = 20.03 \pm 0.30$ cm$^{-2}$. 
This value encompasses the best-fit $N_\mathrm{HI}$ for all cases where $\overline{x}_\mathrm{HI} \leq 50\%$.
Moreover, the values of the $\overline{x}_\mathrm{HI} > 50\%$ seem unlikely because the best-fit damping wing profiles systematically underestimate the observed fluxes around $\lambda=$~1216--1225~\AA.
Selection of this model was also motivated by the fact that IGM with $\overline{x}_\mathrm{HI} \gtrsim 0.4$ has only been found at higher redshifts \citep[e.g.,][]{2018ApJ...864..142D,2020ApJ...896...23W}. In addition, \cite{2020ApJ...904...26Y} also derive a lower limit for the neutral fraction at $z\sim6$, which is $\overline{x}_\mathrm{HI}\gtrsim10^{-4}$. 
This value was inferred based on the measurements of \Lya\ effective optical depth and hydrodynamical simulations assuming a uniform ultraviolet background.
At the same time, \cite{2020ApJ...904...26Y} also note that their model does not rule out a possibility of $\overline{x}_\mathrm{HI} \sim 0.1$--$0.2$ but not much beyond that.

For completeness, we also attempt to model the \Lya\ damping wing using MUSE and the lower-quality FIRE spectra presented in Appendix~\ref{sec:appendix_d}. The PCA model, in this case, is more strongly impacted by telluric line residuals and less by intrinsic features aside from the general quasar spectral slope. This results in spurious, much higher prediction of \Lya\ line flux, in turn requiring an unphysical neutral fraction of $\overline{x}_\mathrm{HI}\geq80\%$. Hence we do not consider FIRE to add robust and trustworthy independent information to this analysis.

\section{Constraints on a Lyman-alpha halo} \label{sec:halo_find}

\begin{figure*}[htb!]
	\centering
	\gridline{\fig{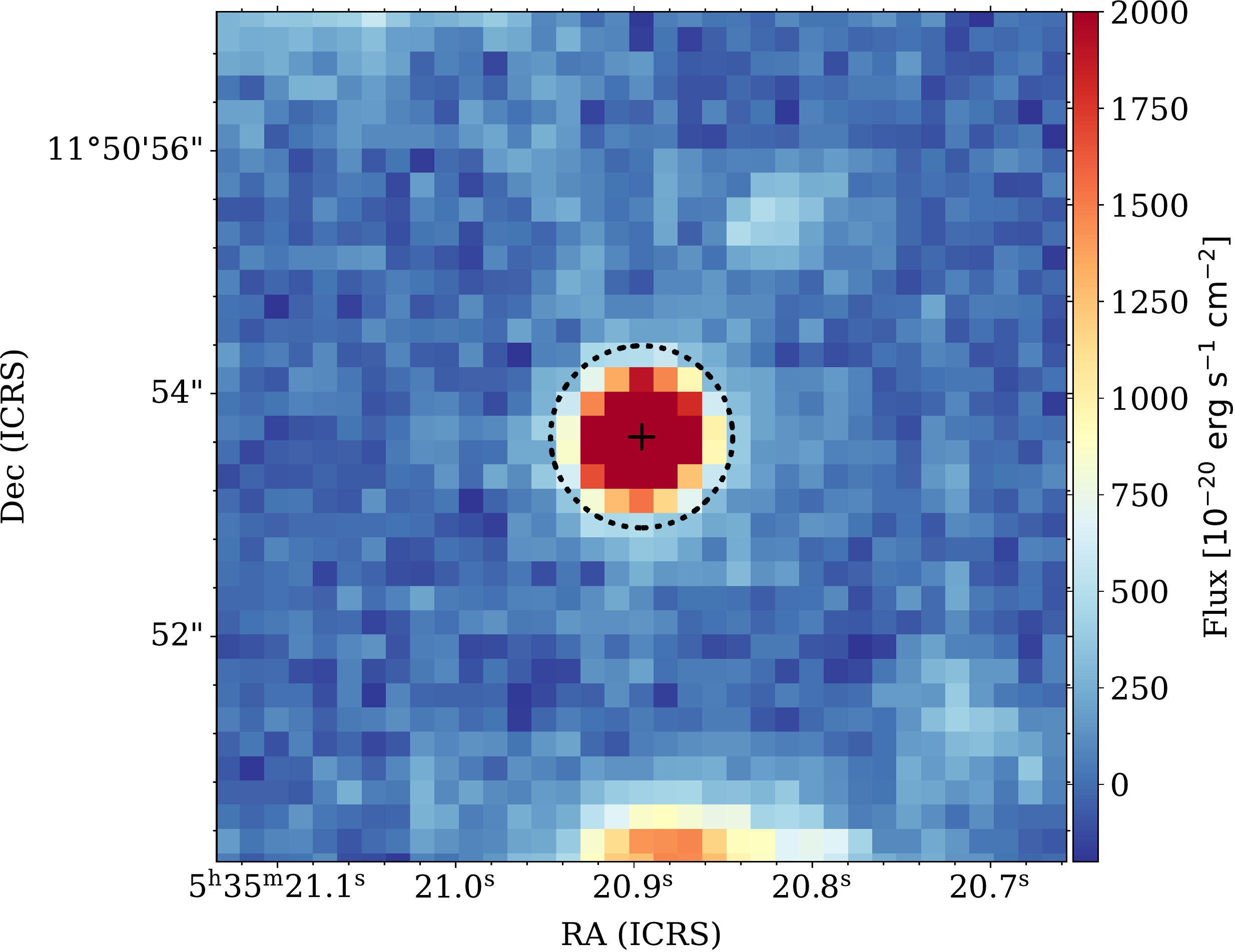}{0.325\textwidth}{Narrow-band \Lya}
		\fig{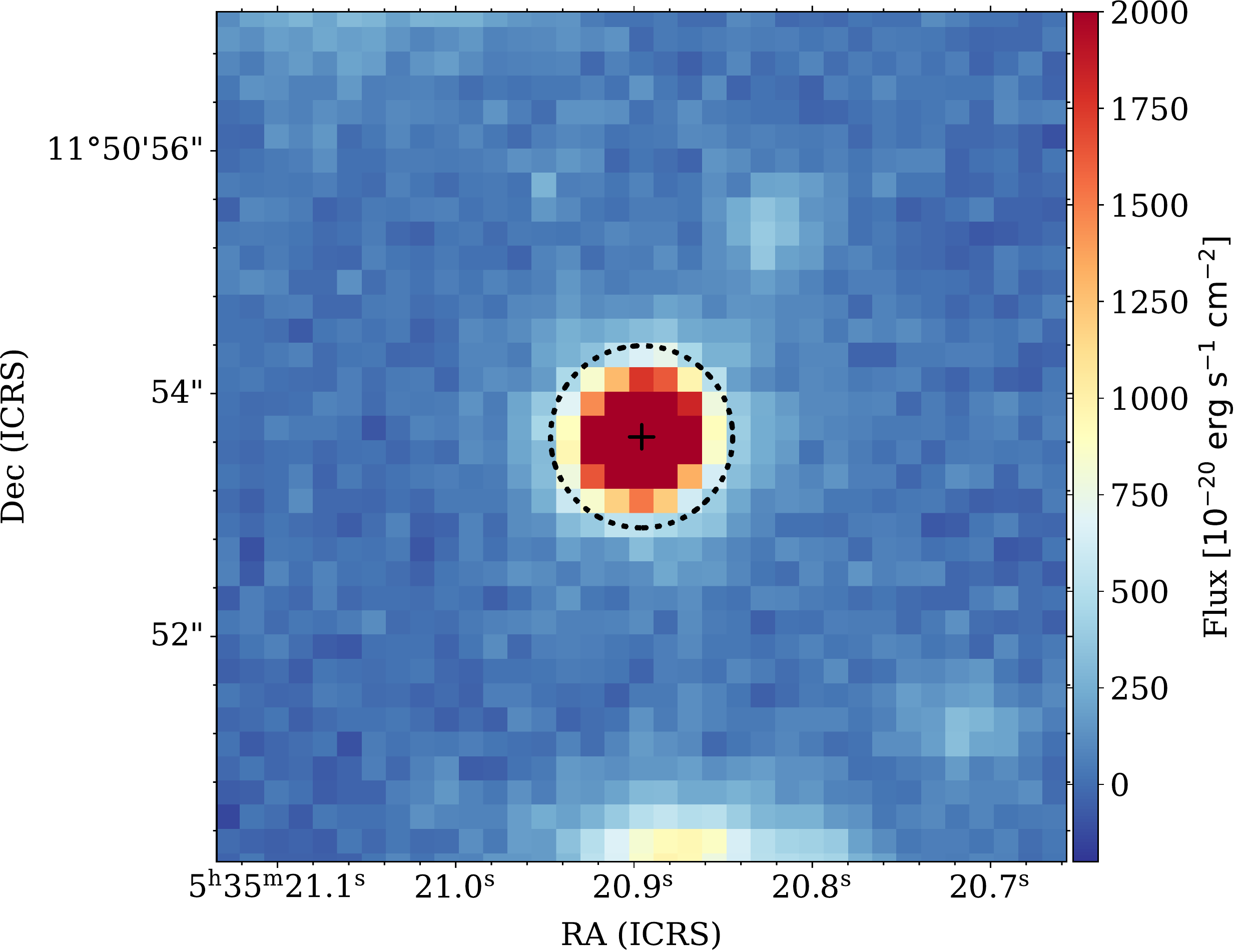}{0.325\textwidth}{Continuum (PSF Model)}
		\fig{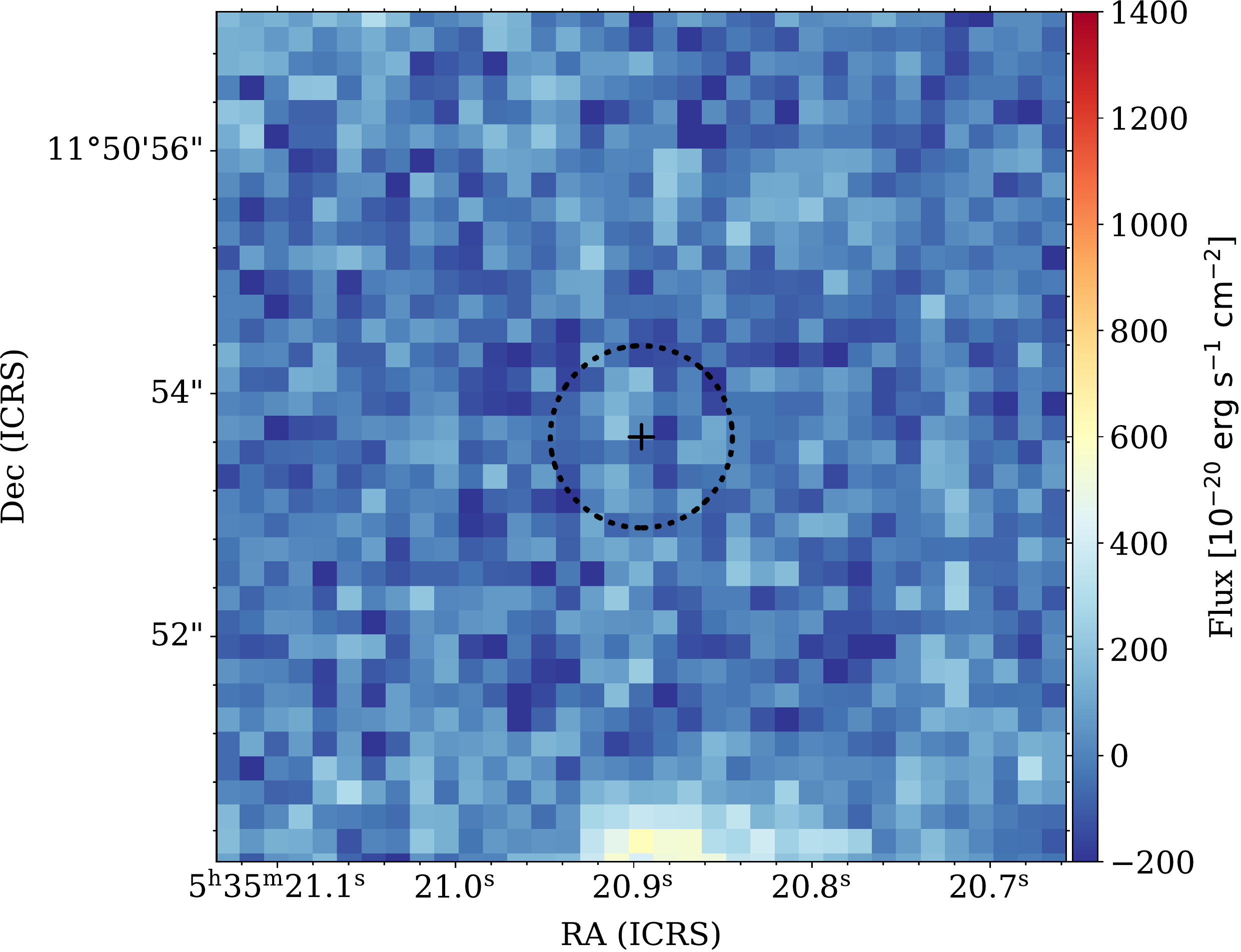}{0.325\textwidth}{Residual Image}
	}
	\caption{
		Pseudo-narrowband images of \PSOJ\ created based on the MUSE data. 
		The black dotted circles mark an aperture with a size of 0\farcs75 used for extracting the quasar spectrum.
		In the \textit{left panel}, we show the pseudo-narrowband image centered on the expected \Lya\ emission wavelength ($\lambda_\mathrm{obs}=$~8850--8997~\AA) as highlighted by the magenta region of the spectrum in the inset of Figure~\ref{fig:spec_model_wide}.
		The continuum, and hence point-source image, constructed by collapsing data cube wavelength layers around the quasar continuum emission, is shown in the \textit{middle panel}. 
		The wavelength range ($\lambda_\mathrm{obs}=$~9100--9200~\AA) combined into this image is denoted in green in the inset of Figure~\ref{fig:spec_model_wide}.
		Finally, the \textit{right panel} shows the pseudo-narrowband image centered on the \Lya\ line after removing the quasar point source with the PSF subtraction procedure explained in Section~\ref{sec:psfmodsub}. 
		There is no obvious detection of \Lya\ halo as seen in the point-source-subtracted residual image.
        As a side note, the bright emission to the south of the central quasar comes from a foreground galaxy located at $z\approx0.29$.
	}
	\label{fig:psfsub_muse}
\end{figure*}

To find alternative support for the young quasar contention, we exploited MUSE data for constraining the extension of the \Lya\ halo around the \PSOJ.
Previously discovered \Lya\ halos around early quasars \citep{2019ApJ...887..196F} guide our expectation on the potential extent ($\lesssim 30$ kpc) and luminosity ($\log L_\mathrm{Ly\alpha} \sim 42 - 44$~erg~s$^{-1}$). 
These parameters set a clear need for a removal of the quasar continuum point-source light, smeared out by the VLT/MUSE $\sim$0\farcs5 PSF, before a search for the halo can be carried out.

\subsection{Quasar Point-source Modeling and Subtraction} \label{sec:psfmodsub}
For all practical purposes, the quasar’s accretion disk can be considered as a point source in the rest-frame ultraviolet and optical.
For any given IFU wavelength, the point-spread function (PSF) can be constructed based on neighboring bright stars in the frame, or -- for regions of line emission -- from adjacent spectral regions of quasar continuum \citep{2017ApJ...848...78F}.

Using \cite{2019ApJ...881..131D} as a reference, we modeled the PSF from data in the following steps.
Several spectral layers in the MUSE cube containing quasar's continuum were chosen and collapsed to produce a local PSF model.
Ideally, we would need to include a spectral range as wide as possible to increase the S/N.
However, there is on one side contamination from night-sky emissions across the wavelength of interest that modify noise properties of the adjacent wavelength layers, resulting in S/N degradation. On the other side, the PSF is chromatic due to wavelength-dependent diffraction in the air of different densities, requiring to choose wavelength ranges for PSF construction close to the wavelength range for which the PSF is constructed. 
Therefore, we collapsed the wavelength layers of the region $\lambda_\mathrm{obs}=$~9100--9200~\AA\ where contamination is minimal, quasar continuum has high enough S/N while staying within a few percent distances to the \Lya\ wavelength (see the inset of Figure~\ref{fig:spec_model_wide}).

With this PSF in hand, we subtract the quasar point source by normalizing the PSF model to match each of the MUSE cube wavelength layers within the \Lya\ spectral region, i.e., at $\lambda_\mathrm{obs}=$~8850--8997~\AA. For this, we simply scale the fluxes measured within an aperture radius of two spatial pixels (0\farcs4, at an angular resolution of 0\farcs3--0\farcs4 for the MUSE data), assuming that this central region is massively dominated by the unresolved quasar's emission \citep[e.g.,][]{2017ApJ...848...78F,2019ApJ...887..196F}.
Finally, subtracting the scaled PSF model cube from the \Lya\ cube layer gave us an entire quasar nucleus-subtracted data cube containing an extended, i.e., not point-source, flux.

\subsection{The Nondetection of a Lyman-alpha Halo}
We show the MUSE-based images of \PSOJ\ in Figure~\ref{fig:psfsub_muse}.
Both \Lya- and continuum-spectrum of the quasar are extracted using a radius of 0\farcs75 as showed by the black dotted circle in the images.
The left panel presents the constructed pseudo-narrowband image centered on the \Lya\ line, the middle image the constructed continuum image we used as a quasar PSF model.
The quasar point-source-subtracted image is shown in the right panel, which is equivalent to the continuum image subtracted from the \Lya\ image on the left.

\begin{figure}[htb!]
	\centering
	\epsscale{1.17}
	\plotone{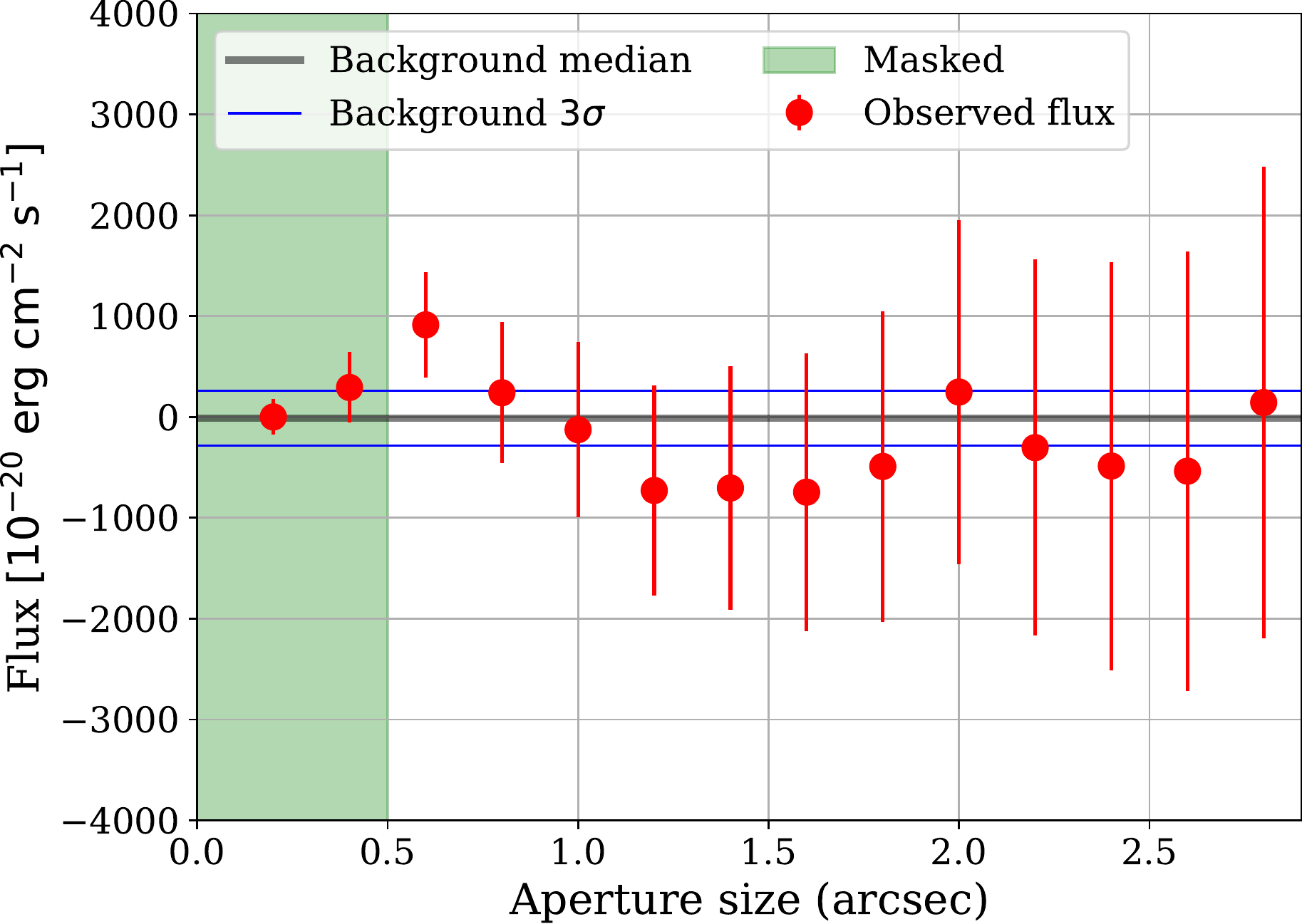}
	\caption{Azimuthally averaged radial light profile of \PSOJ's \Lya\ emission measured from the PSF-subtracted pseudo-narrowband image.
    Tentative detection of the halo flux within a 0\farcs6 aperture radius is observed, but there are no apparent extended \Lya\ emissions beyond this radius.
    }
	\label{fig:halo_flux}
\end{figure}

The image itself does not show an obvious extended halo of high surface brightness. To test this down to fainter surface brightness levels, we followed \citet{2019ApJ...881..131D} and performed aperture photometry with multiple radii to construct an azimuthally averaged radial light profile of potential extended \Lya\ emission (Figure~\ref{fig:halo_flux}).
The estimated $1\sigma$ surface brightness limit of the quasar continuum-subtracted pseudo-narrowband image is $2.76 \times 10^{-18}~\mathrm{erg~s^{-1}~cm^{-2}~arcsec^{-2}}$, corresponding to a \Lya\ luminosity upper limit of $\log L_\mathrm{Ly\alpha} \lesssim 43.46$~erg~s$^{-1}$ at 1\arcsec~aperture radius.
In addition, we also found a marginal halo flux detection within a 0\farcs6 aperture radius, but there are no evident extended \Lya\ emissions beyond this radius.

Previous studies discovered that the \Lya\ halo luminosity does not depend directly on the quasar's instantaneous ionizing flux, which can be probed with the absolute magnitude at 1450~\AA\ \citep[$M_{1450}$; e.g.,][]{2019MNRAS.482.3162A,2019ApJ...881..131D}.
In other words, circumgalactic medium properties, including the ionization state, temperature, and density, play a more important role in producing the \Lya\ halo emission \citep[see, e.g.,][]{2021MNRAS.502..494M}.
Out of the 31 quasars studied by \cite{2019ApJ...887..196F}, 19 of them have absolute magnitudes comparable to \PSOJ, i.e.,  $-26 \lesssim M_{1450} \lesssim 27$.
To make a direct comparison with the depth of \cite{2019ApJ...887..196F} data, we estimate the surface brightness limit of our MUSE image following their prescription.
The calculation is done by collapsing five wavelength channels -- or equivalent to a total of 6.25 \AA\ -- around the expected location of \PSOJ's \Lya\ emission. 
Subsequently, we obtain a $5\sigma$ surface brightness limit of $0.29 \times 10^{-17}~\mathrm{erg~s^{-1}~cm^{-2}~arcsec^{-2}}$ for a circular aperture with a radius of 1\arcsec.
This means that our data are comparable with the depth of their samples -- i.e., surface brightness limits of ranging from 0.1 to $1.1 \times 10^{-17}~\mathrm{erg~s^{-1}~cm^{-2}~arcsec^{-2}}$.

Intriguingly, of the 19 quasars in the \cite{2019ApJ...887..196F} subsample, 10 show halos with luminosities ranging from $\log \mathrm{Ly\alpha} = 42.9$ to 44.3~erg~s$^{-1}$ while the other 9 quasars do not show any signs of extended \Lya\ emission.
In other words, at the current depth of our MUSE data, there is a $\sim50$\% chance that the \PSOJ's \Lya\ halo is intrinsically nonexistent and a 50\% probability that the halo exists but is below our detection limit.
However, if we wanted to decrease the current surface brightness limit by a factor of 2, around 10~hr of additional observing time with MUSE will be needed -- both for \PSOJ\ as well as each of the nine comparison quasars from \citeauthor{2019ApJ...887..196F}
We also note that there are two notably fainter quasars, J2329--0301 ($M_{1450} = -25.19$) and J2228+0110 ($M_{1450} = -24.47$), which show halos with luminosity of at least $\log \mathrm{Ly\alpha} > 43.7$~erg~s$^{-1}$ \citep{2019ApJ...887..196F}.
It is also interesting to note that none of the three known young quasars with estimated lifetimes of only $t_\mathrm{Q} \sim 10^3$--$10^4$~yr -- i.e., J2229+1457 ($M_{1450} = -24.72$), J0100+2802 ($M_{1450} = -29.09$), and J2100--1715 ($M_{1450} = -25.50$) -- show presence of extended \Lya\ halo \citep{2019ApJ...887..196F,2020ApJ...904L..32D,2021ApJ...917...38E}.
In the case of \PSOJ, this might indicate that there is not enough \Lya\ diffuse gas surrounding quasar to emit a halo, or, alternatively, ionizing light from the young quasar phase has not had time to travel far enough.
Note that the light travel time for a halo with a size of 0\farcs6, or equivalent to 3.32~kpc, corresponds to $\sim10^4$~yr.

\section{Discussion} 
\label{sec:discussion}

\subsection{Elemental Abundance Ratios}

The new $z=6.314$ proximate absorber presented here resembles the low-ionization systems characterized by \citet{2019ApJ...882...77C} and \citet{2020arXiv201110582S}, where large column densities of elements like \mgii, \cii, and \oi\ are observed but we find no absorptions from highly ionized elements such as \ion{Si}{4} or \civ.
In our case, we found $\log N_\mathrm{MgII} = 13.33 \pm 0.03~\mathrm{cm^{-2}}$, $\log N_\mathrm{CII} = 14.23 \pm 0.09~\mathrm{cm^{-2}}$, $\log N_\mathrm{OI} = 14.53\pm0.32~\mathrm{cm^{-2}}$,
$\log N_\mathrm{SiIV} < 12.91~\mathrm{cm^{-2}}$, and $\log N_\mathrm{CIV} < 13.87~\mathrm{cm^{-2}}$.
This source also differs from the typical lower-redshift DLAs ($2 \lesssim z \lesssim 4$), which are likely to have the associated \civ, but, for the given redshift, this is not unexpected due to a decreasing rate of high-ionization absorbers at $z\gtrsim5$ \citep{2019ApJ...882...77C}.
On the other hand, the nondetection of \mgi\ with $\log N_\mathrm{MgI} < 11.98~\mathrm{cm^{-2}}$ implies some degree of ionization from the radiation field of massive stars in the far-ultraviolet regime \citep{2020arXiv201110582S}.
In addition, we also could not find other associated metal line absorptions like \siii, \alii, and \feii\ -- i.e., $\log N_\mathrm{SiII} < 13.16~\mathrm{cm^{-2}}$, $\log N_\mathrm{AlII} < 12.15~\mathrm{cm^{-2}}$, and $\log N_\mathrm{FeII} < 12.67~\mathrm{cm^{-2}}$ -- that we would expect for a specifically strong DLA system.

We report the element ratios relative to the solar abundances \citep{2009ARA&A..47..481A} in Table~\ref{tab:metallicity}.
For O and C, we derive the solar abundances based on the photospheric values while for other elements we utilize the meteoritic values \citep[see, e.g.,][]{2019ApJ...885...59B}.
Note that the observed $\log N_\mathrm{HI} = 20.03 \pm 0.30$ cm$^{-2}$ means that this cloud is more similar to a sub-DLA system \citep[$19.0 < \log N_\mathrm{HI} < 20.3$~cm$^{-2}$;][]{2010MNRAS.408.2071M,2019ApJ...882...77C} and the actual elemental abundances might need ionization and dust depletion corrections with a total factor of $\sim 0.0$--0.7~dex. \citep[e.g.,][]{2010MNRAS.408.2071M,2016MNRAS.458.4074Q,2021MNRAS.502.4009B}.

\begin{deluxetable}{cccc}
	\tablecaption{Elemental abundances of the $z=6.314$ sub-DLA in the line of sight of \PSOJ.}
	\label{tab:metallicity}
	\tablehead{
		\colhead{X} & \colhead{$\log \epsilon(\mathrm{X})_\odot$\tablenotemark{a}} & \colhead{$\log N_\mathrm{X}$ (cm$^{-2}$)} & \colhead{[X/H]}
	}	
	\startdata
	H & 12.00 & $20.03 \pm 0.30$ & -- \\
	C & 8.43 & $14.23 \pm 0.09$ & $-2.23 \pm 0.31$ \\
	O & 8.69 & $14.53 \pm 0.32$ & $-2.19 \pm 0.44$ \\
	Mg & 7.53 & $13.33 \pm 0.03$ & $-2.23 \pm 0.30$ \\
	Al & 6.43 & $<12.15$ & $<-2.31$ \\
	Si & 7.51 & $<13.16$ & $<-2.38$ \\
	Fe & 7.45 & $<12.67$ & $<-2.81$ \\
	\enddata
	\tablecomments{The calculated relative abundances do not include dust depletion and ionization corrections. The elemental ratios are relative to the solar abundances, i.e., $[\mathrm{X/Y}] = \log(N_\mathrm{X}/N_\mathrm{Y}) - \log(N_\mathrm{X}/N_\mathrm{Y})_\odot$.}
	\tablenotetext{a}{The definition is $\log \epsilon(\mathrm{X})_\odot = 12 + \log(N_\mathrm{X}/N_\mathrm{H})_\odot$.}
\end{deluxetable}

\begin{figure}[htb!]
	\centering
	\epsscale{1.17}
	\plotone{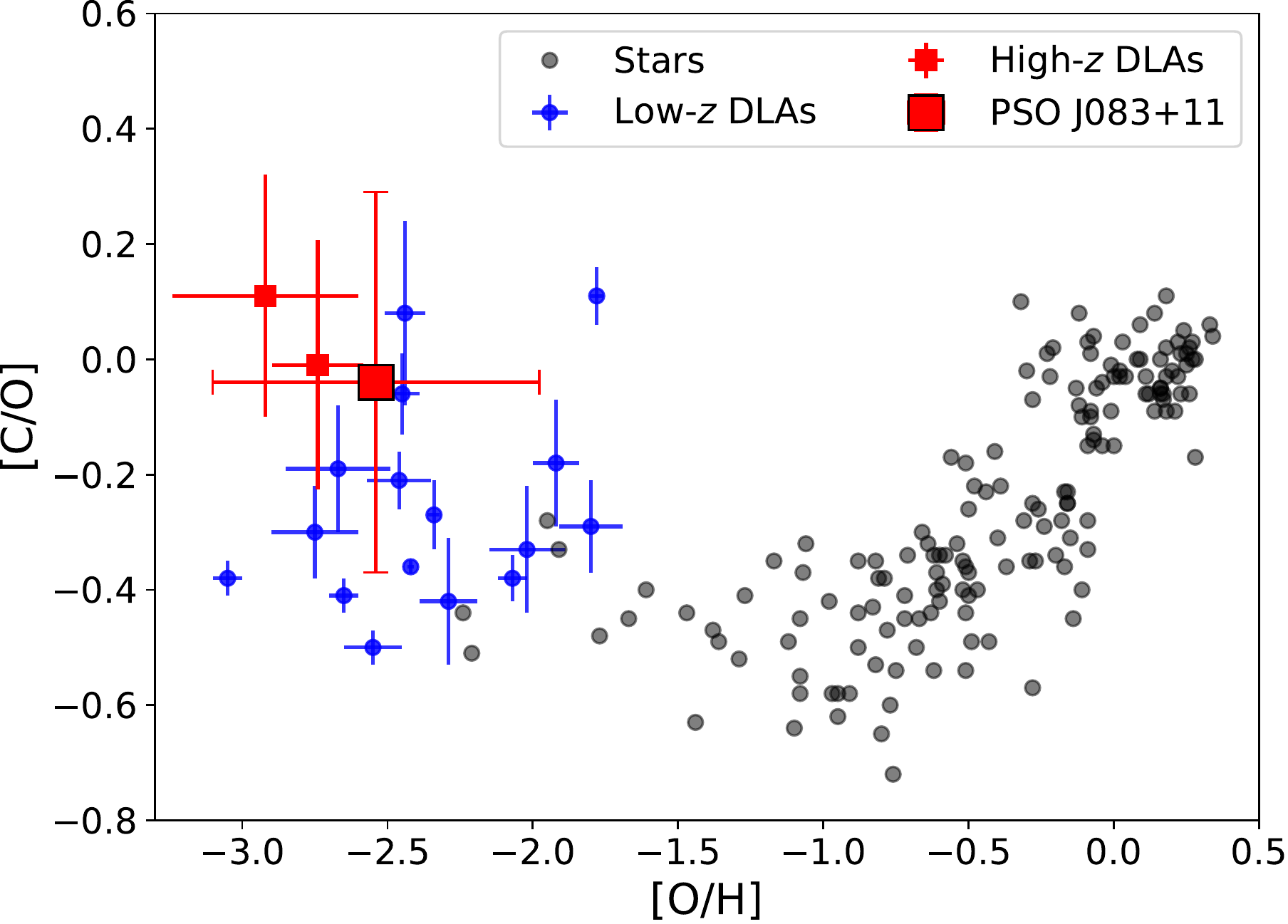}
	\caption{
	Relation between metallicity [O/H] and chemical evolution of [C/O]. 
	Measurements for Milky Way stars taken from \citet{2009A&A...500.1143F} and \citet{2014A&A...568A..25N} are marked as gray circles, while low-redshift DLAs ($2 \lesssim z \lesssim 4$) from \citet{2017MNRAS.467..802C} are denoted with blue circles. 
	The $z\gtrsim6$ metal-poor DLAs from \citet{2019ApJ...885...59B} and \citet{2018ApJ...863L..29D} are shown with red squares.
	Specific for the sub-DLA toward \PSOJ, we show a corrected metallicity where we adopt a correction factor of 0.35$\pm$0.35~dex (see text) to account for ionization and depletion effects.
	Consequently, this changes the measured [O/H] from $-2.19 \pm 0.44$ to $-2.54 \pm 0.56$.}
	\label{fig:chemical_evolution}
\end{figure}

There is a known stellar relation that the [C/O] element ratio is linearly increasing  with metallicity in the range $-0.5 < \mathrm{[C/O] < 0.5}$ and [O/H]~$>-1$.
An explanation is an increase in carbon production by rotating massive stars -- which also increases with metallicity -- plus a delayed carbon production from stars with lower masses \citep{2004A&A...414..931A}.
On the other hand, at [O/H]~$\lesssim-1$, it there is an opposite trend observed in metal-poor DLAs \citep{2017MNRAS.467..802C,2019ApJ...885...59B} and metal-poor stars \citep{2004A&A...414..931A,2009A&A...500.1143F} -- i.e., an actual increases in [C/O].
\PSOJ\ seems to follow this empirical trend, showing similar [C/O] abundance and [O/H] metallicity compared to other metal-poor DLAs (see the red square in Figure~\ref{fig:chemical_evolution}).
The calculated abundance is [C/O]~$=-0.04 \pm 0.33$ while the derived metallicity is [O/H]~$=-2.19 \pm 0.44$.
We refer the reader to \citet{2011MNRAS.417.1534C} and \citet{2017MNRAS.472.3532M} for further references on the possible formation scenarios that could clarify the chemical composition of the DLAs.

\subsection{Is the Presence of a DLA Truncating PSO J083+11 Proximity Zone?}

Around the end of the reionization epoch ($z\sim6$) there are regions of the IGM with partially still neutral hydrogen components, suppressing all photon transmission blueward of the \Lya\ wavelength \citep[e.g.][]{1965ApJ...142.1633G}.
However, a luminous quasar might be capable of ionizing the adjacent medium with its intense radiation, producing a bubble of enhanced transmission in the nearby \Lya\ forest titled as the proximity zone \citep[e.g.,][]{2006AJ....132..117F,2017ApJ...840...24E}.

In \citet{2020ApJ...903...34A}, we argued that \PSOJ\ is having a small size of proximity zone in absolute terms ($R_\mathrm{p} = 1.17\pm 0.32$~Mpc) due to a limited lifetime of unobscured accretion, and hence unobscured emission into its environment ($t_\mathrm{Q} = 10^{3.4\pm0.7}$~yr; i.e., a young quasar).
This measured proximity zone is substantially below the size for typical $z\sim6$ quasars with similar luminosity as \PSOJ, i.e., $R_\mathrm{p, typical} \sim 4$~Mpc (\citealt{2020ApJ...900...37E}, see their Figure~7).
Correspondingly, the accretion lifetime itself can be inferred from the quasar's proximity zone size, where smaller sizes correspond to a younger lifetime \citep{2018ApJ...867...30E,2020MNRAS.493.1330D,2020ApJ...900...37E,2021ApJ...911...60C}.
However, due to the limitation of our earlier initial data, we could not constrain whether or not the proximate absorber might play a significant role in blocking ionizing radiation from the central quasar to the IGM -- potentially making the proximity zone measurement inaccurate.

We now improved on this after locating and characterizing the proximate sub-DLA system at $z=6.314$ with a column density of neutral hydrogen of $\log N_\mathrm{HI} = 20.03 \pm 0.30$ cm$^{-2}$.
In Section~\ref{sec:damping_fit}, we stepped over a range of IGM damping wing shapes using a grid of constant $\overline{x}_\mathrm{HI}$ and proximity zone sizes, then fitted a Voigt profile as the sub-DLA absorption model to the already attenuated continuum.
From this exercise, we can only obtain the lower limit of the proximity zone size, i.e., $R_\mathrm{p} \gtrsim 1.35$~Mpc, which corresponds to a quasar lifetime of $t_\mathrm{Q} \gtrsim 10^{3.5}$~yr.
We caution that, if the ionization bubble created by the quasar reaches the sub-DLA location and beyond but is hidden from the observer due to the blockage by the sub-DLA, the proximity zone size will no longer be sensitive to quasar age.
Therefore, another independent diagnostic for a young quasar lifetime would be required.

We find such evidence based on the observed \Lya\ halo size of \PSOJ\ for this case.
As a rough calculation, we estimate the size as the distance traveled by the light from the central accretion disk to the halo as $R_\mathrm{halo} = c \times t$. Here, $c$ is the light speed, and $t$ is the light travel time.
In Figure~\ref{fig:halo_flux} we see the detection of a halo within a 0\farcs6 aperture radius, but then the fluxes are decreasing and hit the background level at around 1\arcsec.
In other words, the halo would have a minimum projected size of 3.32~kpc with a maximum radial extent up to 5.54~kpc, and there is no detection of extended \Lya\ emissions beyond this radius.
For a halo with that size, this corresponds to a light travel time of $t \sim 10^4$~yr, providing additional and independent evidence supporting the young quasar scenario for \PSOJ.
If the quasar has been accreting for a much longer period (i.e., $t_\mathrm{Q}\gtrsim10^5$~yr), but its ionizing radiation could not be observed due to obscuration along our sightline, the \Lya\ halo around the quasar is likely to extend along the other unobscured sightlines \citep{2019ApJ...887..196F,2021ApJ...917...38E}.
For \PSOJ, the aforementioned case seems unlikely because we do not find any sign of extended \Lya\ emission.
In contrast, the \Lya\ ionized nebula is predicted to be modest or absent if the quasar's radiation has only recently turned on \citep{2018ApJ...864...53E}.

As a side note, we attempted to look for the galaxy's emission associated with the \PSOJ's sub-DLA using ALMA data (see Appendix~\ref{sec:appendix_b} for details).
A neighboring [\cii]\,158~$\mu$m emitter, J083.8372+11.8474, at $z=6.3309$ is detected in the southwest direction from \PSOJ, where the projected separation between them is 2\farcs88 ($\sim 16$~kpc).
This source has an integrated [\cii] flux of $0.61\pm0.01$~Jy~km~s$^{-1}$, FWHM of $318\pm58$~km~s$^{-1}$, and luminosity of $L_\mathrm{[CII]} = 6.22\pm0.14 \times 10^{8} L_\odot$.
Its [\cii] line properties are similar to those of $z\sim4$ DLAs studied by \cite{2019ApJ...870L..19N} as well as companion galaxies around $z\gtrsim6$ quasars \citep[e.g.,][]{2019ApJ...882...10N,2021A&A...652A..66P}.
However, the [\cii]-based redshift of this \PSOJ's companion galaxy significantly differs from the sub-DLA's redshift estimated via the centroids of rest-frame ultraviolet metal absorption lines.
The velocity offset of $\Delta V=692$~km~s$^{-1}$ between those two sources might indicate that they are two unrelated galaxies.

\section{Summary and Conclusion} 
\label{sec:summary}
In this study, we characterized the environment and absorption systems toward \PSOJ, a weak-line quasar at $z=6.3401$. 
The strong \Lya\ absorption along with several metal lines (e.g., \mgii, \oi, and \cii) is observed in the Gemini/GNIRS, Magellan/FIRE, and VLT/MUSE spectra.
This indicates the presence of a sub-DLA system at $z=6.314$ and a strong \mgii\ absorber at $z=2.2305$.
To explain the detected \Lya\ damping wing, we model the corresponding absorption profile with a combination of a sub-DLA with a column density of neutral hydrogen of $\log N_\mathrm{HI} = 20.03 \pm 0.30~\mathrm{cm^{-2}}$ plus absorption from an IGM with a neutral fraction of around 10\%.
The sub-DLA toward \PSOJ\ has an abundance ratio of [C/O]~$=-0.04 \pm 0.33$ and metallicity of [O/H]~$=-2.19 \pm 0.44$, similar to those of low-redshift metal-poor DLAs.
The presence of this sub-DLA truncates \PSOJ's proximity zone and complicates the quasar lifetime measurement.
However, at the same time, this quasar shows no sign of \Lya\ halo, where the estimated $1\sigma$ surface brightness limit is $2.76 \times 10^{-18}~\mathrm{erg~s^{-1}~cm^{-2}~arcsec^{-2}}$ at 1\arcsec\ aperture radius, or corresponds to \Lya\ luminosity of $\leq 43.46$~erg~s$^{-1}$.
This nondetection provides an alternative and independent support for the young quasar hypothesis, where the unobscured accretion lifetime leads to small or no \Lya\ halo simply from a so-far limited light travel distance.

To obtain more evidence in the future, observations of extended narrow line emissions around \PSOJ\ might be needed.
If this object is really in the early stage of quasar activation, the radial size of the extended narrow line region -- which can be traced with [\ion{O}{3}]~$\lambda5007$ or H$\alpha$~$\lambda6563$ -- would have to be small.
This would require high-sensitivity, high-angular-resolution mid-infrared observations with the James Webb Space Telescope.
Building up a more coherent sample of young quasars with high black hole masses at this epoch will be the next step to understanding whether these are intermittent accretion phases in late-stage SMBH formation -- or whether we are reaching the limit of observing the bulk of SMBH mass buildup through optical--NIR selection techniques, if most of SMBH occurs through obscured, possibly radiatively inefficient accretion modes.

As a very different aspect, one way to understand more about the epoch of reionization and the formation of the first galaxies is by studying the absorption systems toward the highest-redshift quasars.
These systems, which at $2 \lesssim z \lesssim 4$ resemble present-day dwarf galaxies \citep{2015ApJ...800...12C}, likely play a significant role in driving the reionization of the IGM, as predicted by numerical simulations.
In just several hundred Myr after the Big Bang, they have experienced substantial enrichment of their gas while retaining a high neutral fraction, which provides an important clue about the contributing stellar populations.
However, direct detections of these dwarfs will be tough even with the next-generation telescope.
Therefore, ``quasar absorption spectroscopy'' is probably still the best way to study them in detail -- and searching for and finding subsequently more and higher redshift DLAs will be a powerful, though challenging, path to fully employ this technique.

\begin{acknowledgments}

We thank the anonymous referees for the constructive comments on the manuscript.
We would like to thank Robert Simcoe for providing the codes for fitting the metal absorption lines.
I.T.A. would personally like to thank Elsa P. Silfia for her incredible support and unending encouragement, particularly in the most difficult times.

This paper is based on observations collected at the European Organisation for Astronomical Research in the Southern Hemisphere under ESO program 0104.B-0665(A).
Part of the data presented in this paper is based on observations obtained at the international Gemini Observatory (GN-2019A-FT-204). Gemini Observatory is managed by the Association of Universities for Research in Astronomy (AURA) under a cooperative agreement with the National Science Foundation on behalf of the Gemini Observatory partnership: the National Science Foundation (United States), National Research Council (Canada), Agencia Nacional de Investigaci\'{o}n y Desarrollo (Chile), Ministerio de Ciencia, Tecnolog\'{i}a e Innovaci\'{o}n (Argentina), Minist\'{e}rio da Ci\^{e}ncia, Tecnologia, Inova\c{c}\~{o}es e Comunica\c{c}\~{o}es (Brazil), and Korea Astronomy and Space Science Institute (Republic of Korea).
This paper includes data gathered with FIRE at 6.5~m Magellan Baade Telescopes located at Las Campanas Observatory.
This paper makes use of the following ALMA data: ADS/JAO.ALMA\#2019.1.01436.S. ALMA is a partnership of ESO (representing its member states), NSF (USA), and NINS (Japan), together with NRC (Canada), MOST and ASIAA (Taiwan), and KASI (Republic of Korea), in cooperation with the Republic of Chile. The Joint ALMA Observatory is operated by ESO, AUI/NRAO, and NAOJ.

\end{acknowledgments}

\vspace{5mm}
\facilities{VLT:Yepun (MUSE), Magellan:Baade (FIRE), Gemini:Gillett (GNIRS)}

\software{
	APLpy \citep{aplpy2012,aplpy2019},
    Astropy \citep{2013A&A...558A..33A,2018AJ....156..123A},
    CosmoCalc \citep{2006PASP..118.1711W},
	Linetools \citep{2016zndo....168270P},
	Lmfit \citep{2019zndo...3381550N},
	Matplotlib \citep{2019zndo...2893252C},
	NumPy \citep{2020Natur.585..357H},
	Pandas \citep{2020zndo...3509134R},
	Photutils \citep{2020zndo....596036B},
	SciPy \citep{2020SciPy-NMeth},
	Spectral-cube \citep{2016ascl.soft09017R},
	SpectRes \citep{2017arXiv170505165C}
}

\clearpage

\appendix

\section{Posterior Distributions of Parameters} \label{sec:appendix_c}

\begin{figure*}[htb!]
	\centering
	\epsscale{1.17}
	\plotone{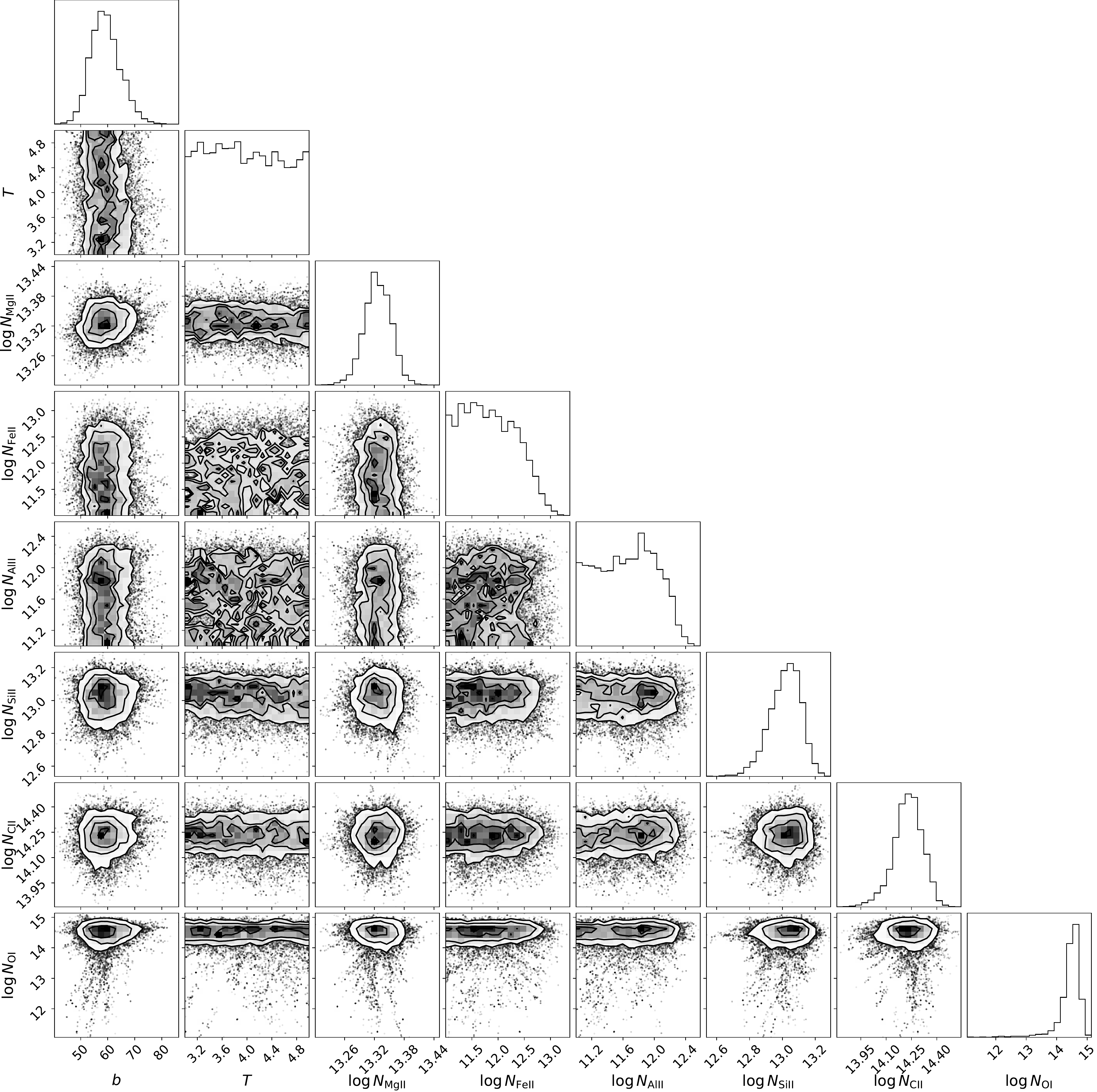}
	\caption{
        Corner plot showing posterior distributions of the thermal parameter ($T$), turbulent parameter ($b$), and metal column densities listed in Table \ref{tab:metallicity}.
        The inferred $b$ seems biased toward large values while $T$ is highly uncertain (see text).
        The column densities of \mgii\ and \cii\ are well constrained because they are clearly detected in the spectrum (see Figure\ref{fig:absline}).
        On the other hand, the marginal detection of \oi\ makes its inferred column density has higher error.
        Nondetections are obvious for \alii\ and \feii\ where their posteriors extend to the minimum value of the input prior.
        We also adopt the upper limit value for \siii\ because its presence is not evident in the spectrum.
	}
	\label{fig:corner_plot}
\end{figure*}

We show the posterior distributions of the thermal parameter ($T$), turbulent parameter ($b$), and metal column densities in Figure~\ref{fig:corner_plot}.
The inferred $b$ seems biased toward large values while $T$ is loosely constrained, probably because we only consider the absorption model with one velocity component.
Metal absorbers at lower $z$ tend to show quiescent kinematics with $b \lesssim 30$~km~s$^{-1}$ \citep[e.g.,][]{2019ApJ...882...77C}.
It might be possible that the sub-DLA toward \PSOJ\ contains clouds with more than one velocity component.
However, this is difficult to constrain using our current data due to spectral resolution limitations.
Nevertheless, \cite{2020arXiv201110582S} found that their code is robust enough for calculating the cumulative column density ($N$) in case of unresolved multiple narrow components, even though the $N$ of the individual clouds is highly uncertain.

\section{Damping Wing Modeling using the MUSE and FIRE Spectrum} \label{sec:appendix_d}

\begin{figure*}[htb!]
	\centering
	\epsscale{1.17}
	\plotone{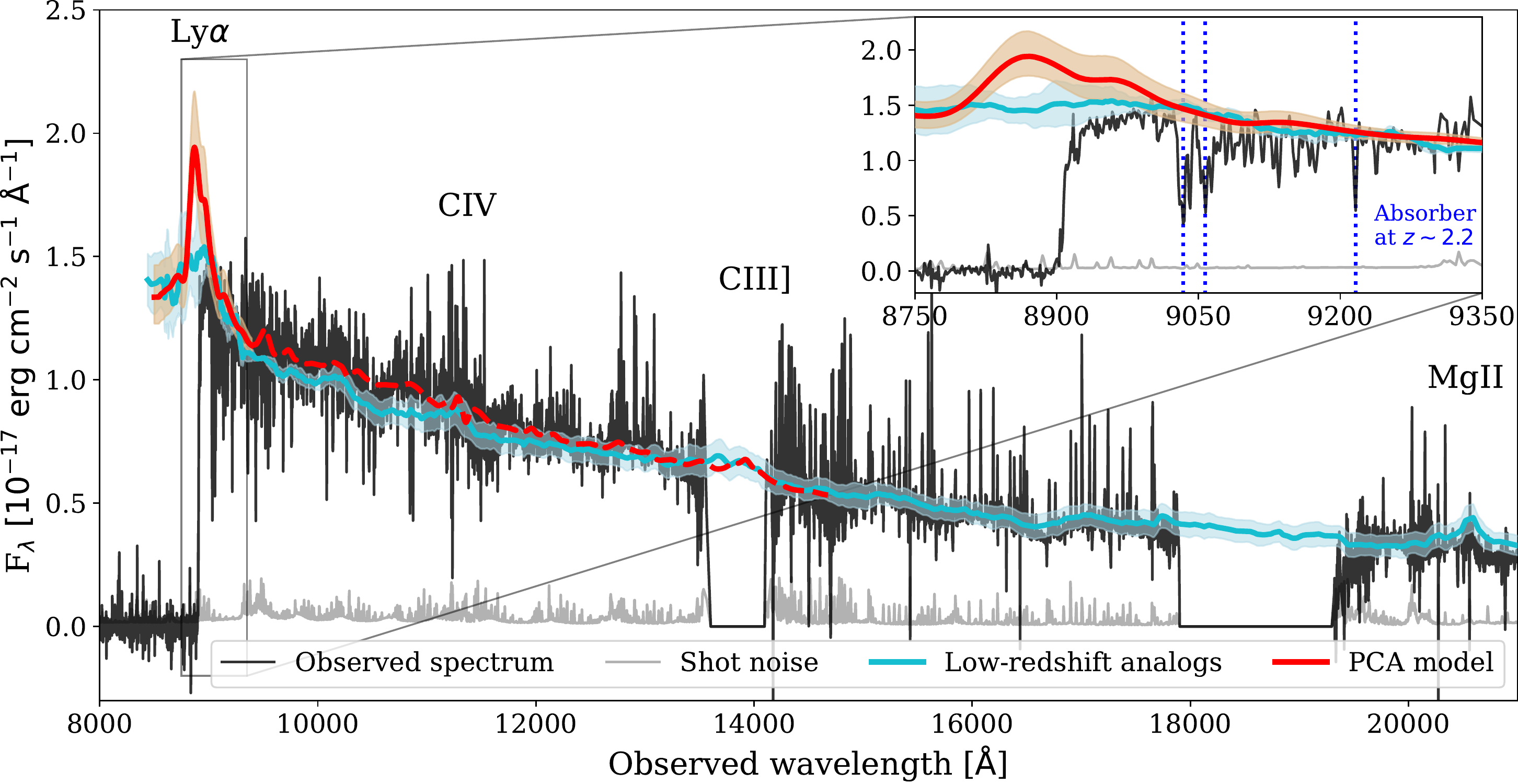}
	\caption{
		MUSE + FIRE spectrum of \PSOJ\ (black line) and associated shot noise (gray line).
		The median composite spectrum of the low-redshift quasar analogs is shown as a cyan line with the $1\sigma$ dispersion around the median as a light blue region.
		The PCA model to predict the blue side of the quasar spectrum and its $1\sigma$ dispersion are denoted with the red line and shaded region, respectively.
		The wavelength range and spectrum used in the PCA fit to predict \Lya\ is shown with the red dashed line.
        The inset figure shows the zoom-in to the region around \Lya.
	}
	\label{fig:spec_model_wide_fire}
\end{figure*}

\begin{figure*}[htb!]
	\centering
	\epsscale{1.17}
	\plottwo{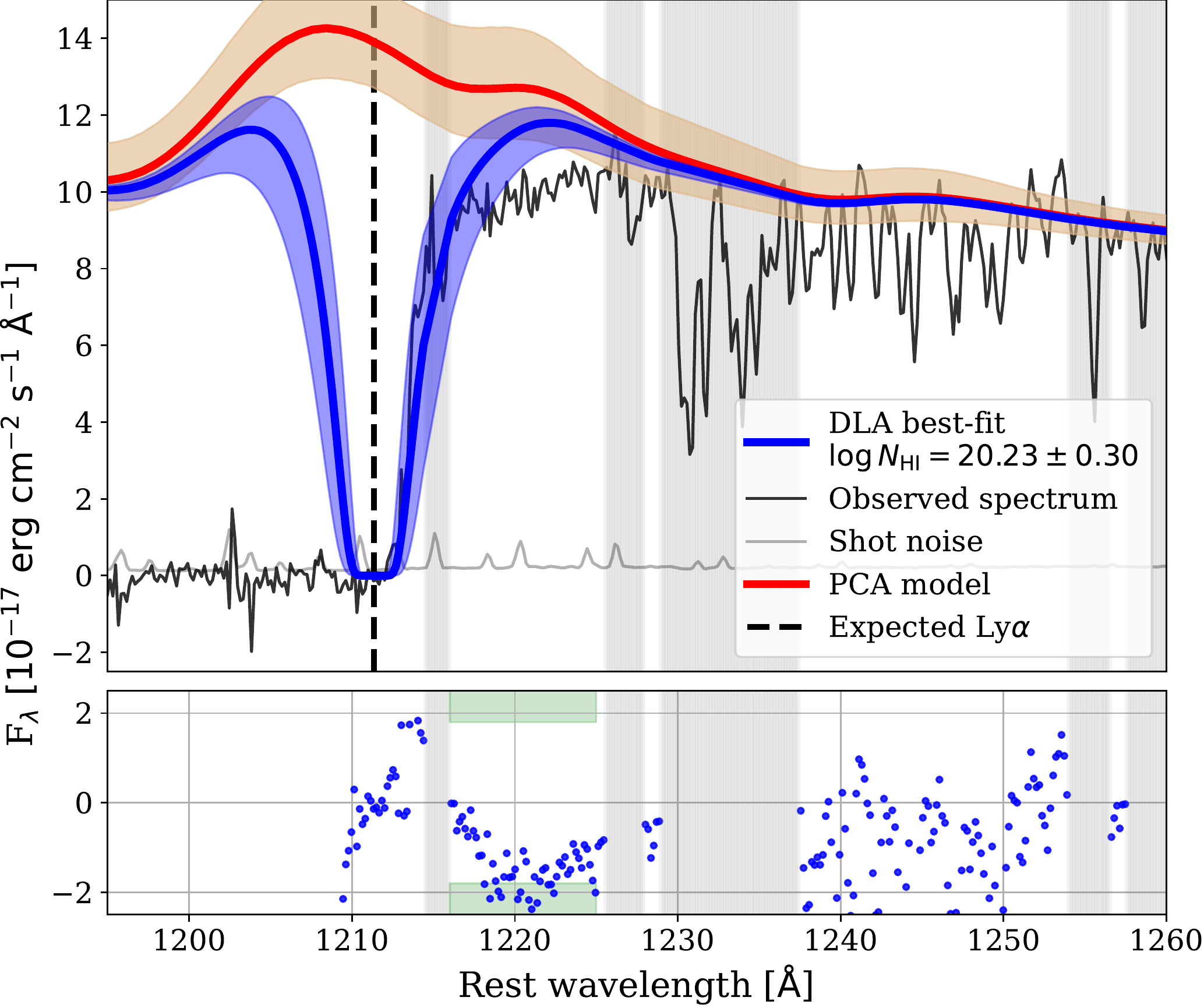}{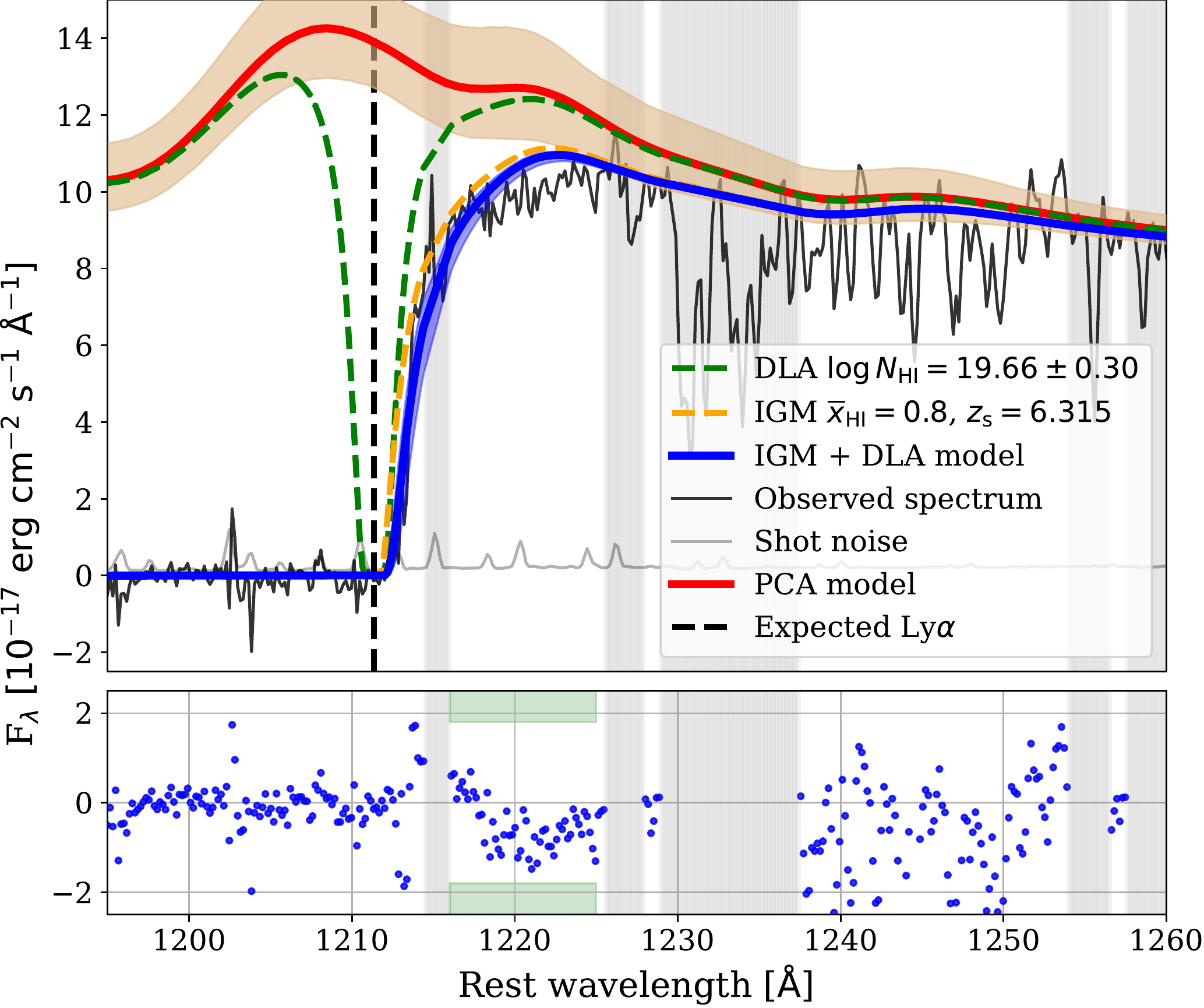}
	\caption{
		MUSE + FIRE spectrum of \PSOJ\ (black line), shot noise (gray line),  and the associated PCA model (red line and shaded region) around the wavelength region where the quasar's \Lya\ emission is expected.
    	The \textit{left panel} shows the absorption from a $z=6.314$ sub-DLA with a hydrogen column density of $\log N_\mathrm{HI} = 20.23 \pm 0.30$ cm$^{-2}$ (blue line and shaded region).
		The \textit{right panel} presents a joint model where we incorporated attenuation due to an IGM with $\overline{x}_\mathrm{HI} = 0.8$ (orange dashed line) and sub-DLA with $\log N_\mathrm{HI} = 19.66 \pm 0.30$ cm$^{-2}$ (green dashed line).
        This joint IGM + DLA model (blue line and shaded region) reconstructs the data at $\lambda = $1216--1225~\AA\ better (see wavelengths marked with green), but it seems to be unphysical (see text). 
        The residual of the fits (blue circles) and masked wavelengths to avoid strong absorption lines or regions with large uncertainty in the sky subtraction (gray shaded area) are also presented in the figure.
	}
	\label{fig:spec_model_fire}
\end{figure*}

As mentioned in Section~\ref{sec:damping_fit}, we prefer to use the MUSE + GNIRS spectrum for modeling the intrinsic quasar spectrum because it has a better overall signal in the spectral data.
Here, we attempt to use the MUSE and lower-quality FIRE spectra for \Lya\ damping wing modeling as a comparison.
\PSOJ's intrinsic emission was very difficult to be modeled using this dataset.
As seen in Figure~\ref{fig:spec_model_wide_fire}, there are spurious spikes in the spectrum -- likely caused by imperfect telluric correction and hence spectrophotometric calibration at each wavelength due to low S/N -- which make it nearly impossible for the PCA to anchor on relevant emission features. 
We had to extensively perturb the PCA fit to get a sensible solution that does not run into unphysical parameters. 
It turns out that PCA fit did not converge to a proper minimum within the physical parameter range. 
Despite the complications in modeling \PSOJ's intrinsic emission, we attempt to fit the \Lya\ absorption as explained in Section~\ref{sec:damping_fit}, using the nominal \Lya\ line prediction by the PCA model. 
These result in best-fit value of $\log N_\mathrm{HI} = 20.23 \pm 0.30$ cm$^{-2}$ for pure sub-DLA model and $\log N_\mathrm{HI} = 19.66 \pm 0.30$ cm$^{-2}$ for the joint IGM + DLA model.
However, the pure sub-DLA model systematically overestimates the fluxes around $\lambda=$~1216--1225~\AA\ (see Figure~\ref{fig:spec_model_fire}).
In addition to that, the joint IGM + DLA model requires an IGM with hydrogen neutral fraction of $\overline{x}_\mathrm{HI} = 0.8$, which seems to be substantially too high at $z\sim6$ and contradicts measurements of prior studies \citep[e.g.,][]{2018ApJ...864..142D,2019ApJ...885...59B,2020ApJ...896...23W,2020ApJ...904...26Y}.
Consequently, we disregard this model to make any further quantitative statements.

\section{Exploring the Effect of Different IGM and Sub-DLA Contributions} \label{sec:appendix_a}

\begin{figure*}[htb!]
	\centering
	\epsscale{1.04}
	\plotone{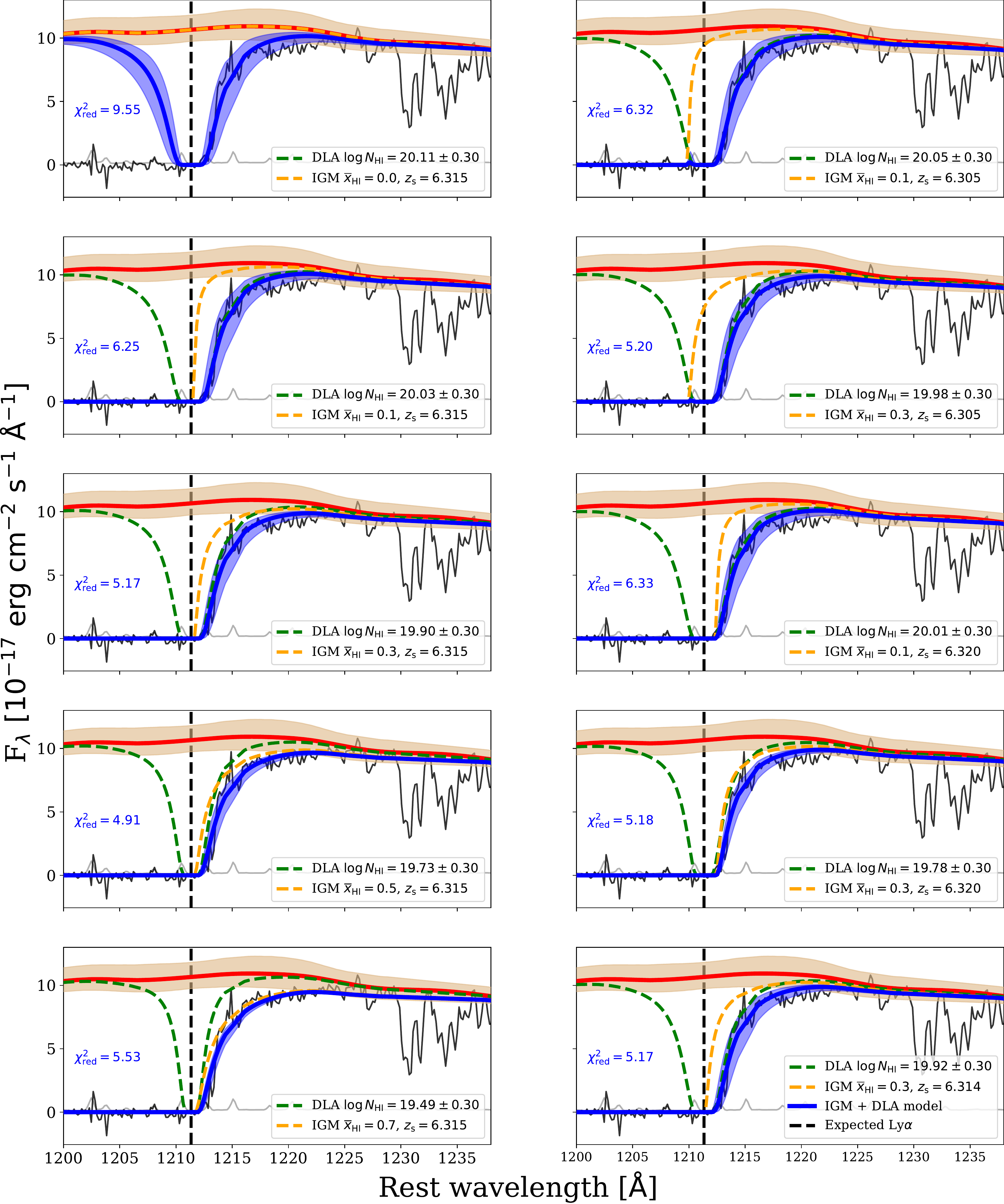}
	\caption{
		\textit{All panels} show the observed spectrum of \PSOJ\ (solid black line), its $1\sigma$ shot noise vector (gray line), PCA continuum model (red line), and best-fit damping wing model (blue line).
		In the \textit{top left panel}, we present the damping wing shape assuming there is zero IGM contribution (i.e., $\overline{x}_\mathrm{HI}=0.0$).
		The rest of the \textit{left panels} present the increasing attenuation by the IGM with the sub-DLA Voigt profiles (green dashed lines; see legends) being modeled using the already attenuated quasar continuum (orange dashed lines) as input. 
		The combination (blue line) is the best-fit match for the observed spectrum (solid black lines).
		The reduced chi-squared ($\chi^2_\mathrm{red}$) value for each IGM + DLA model is indicated with the blue text.
		As discussed in Section~\ref{sec:damping_fit}, the fiducial model that we choose is a 10\% neutral IGM and correspondingly a best-fit sub-DLA profile of $\log N_\mathrm{HI} = 20.03 \pm 0.30$ cm$^{-2}$.
		The \textit{right panels} show the example cases where the quasar's proximity zones extend beyond or within the sub-DLA redshifts but still give comparably good fits.
	}
	\label{fig:joint_fit}
\end{figure*}

In the main text (Section~\ref{sec:damping_fit}), we discussed how the profile of the damping wing in the \PSOJ\ spectrum could be produced by an accumulation of a proximate sub-DLA and a neutral intervening IGM.
Here we explore how different combinations of those two parameters affect the quasar spectrum.
In Figure~\ref{fig:joint_fit} (see left panels) we present cases with neutral hydrogen fractions of $\overline{x}_\mathrm{HI}=$~0.0 to 0.7 and how these values change the best-fit inferred  neutral hydrogen column density of the sub-DLA from $\log N_\mathrm{HI}=$~20.11 to 19.49~cm$^{-2}$.
We also investigate the effect of the quasar's proximity zone to the derived $\overline{x}_\mathrm{HI}$ and $N_\mathrm{HI}$, if it extends beyond certain redshifts (see right panels).
Following the formalism of \citet{1998ApJ...501...15M}, we tried to model the IGM damping wing presuming a constant neutral fraction from the quasar's proximity zone at redshift $z = z_\mathrm{s}$ to $z=5.5$, while being entirely ionized around $z\lesssim5.5$.
We find that values of the IGM neutral fraction $\overline{x}_\mathrm{HI} > 0.5$ seem implausible because the best-fit damping wing profiles systematically do not match the observed fluxes around $\lambda=$~1216--1225~\AA.
On the other hand, the cases with $\overline{x}_\mathrm{HI} \lesssim 0.5$ seem to produce comparably good fits to the data.
In the end, we choose our preferred model as discussed in Section~\ref{sec:damping_fit}, i.e., a 10\% neutral IGM and correspondingly a best-fit sub-DLA profile of $\log N_\mathrm{HI} = 20.03 \pm 0.30$ cm$^{-2}$.
This value covers the best-fit $N_\mathrm{HI}$ values for all cases with $\overline{x}_\mathrm{HI} \leq 50\%$.

\clearpage

\section{Searching for the Sub-DLA Host Galaxy's Emission} \label{sec:appendix_b}

\begin{figure*}[htb!]
	\centering
	\epsscale{1.17}
	\plottwo{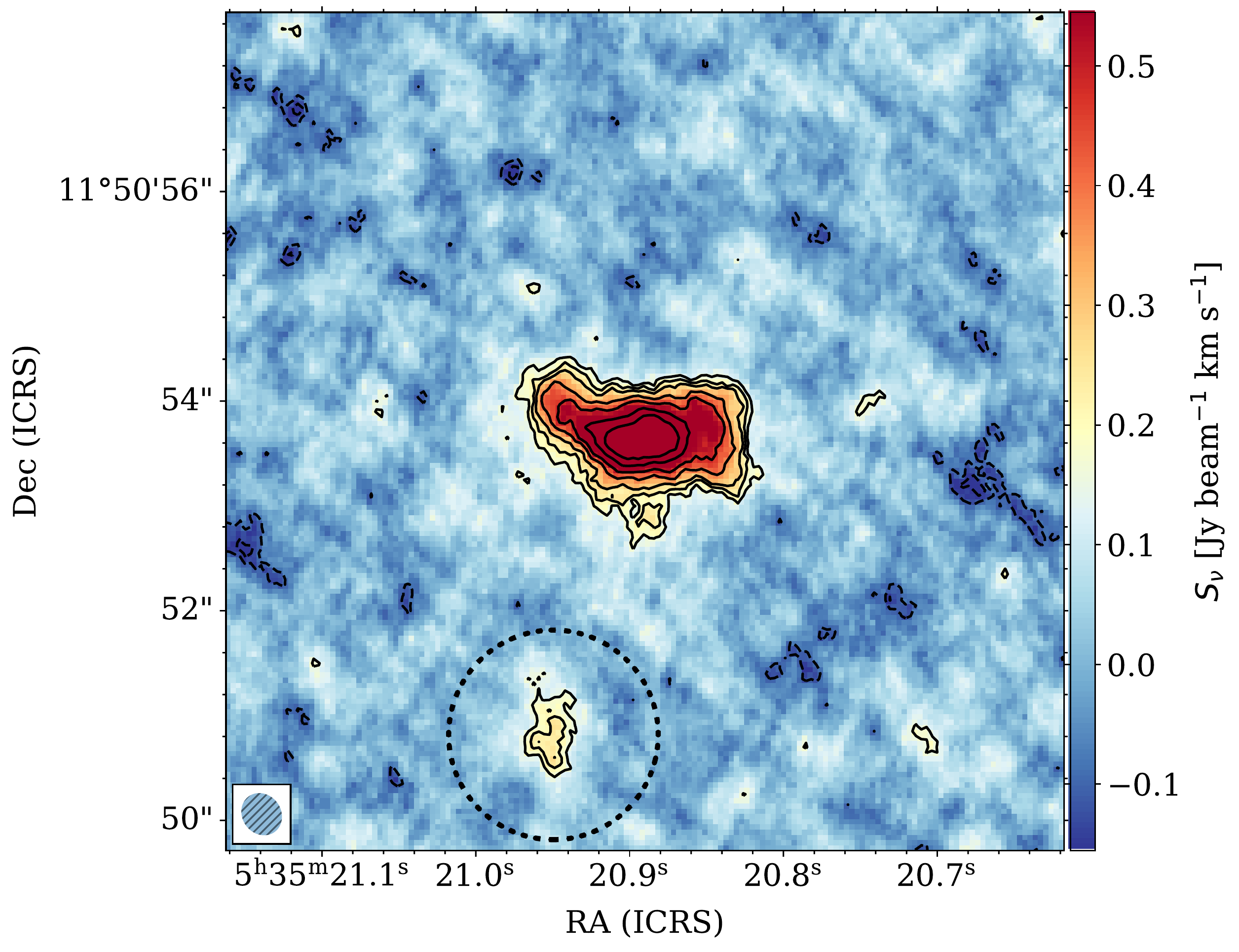}{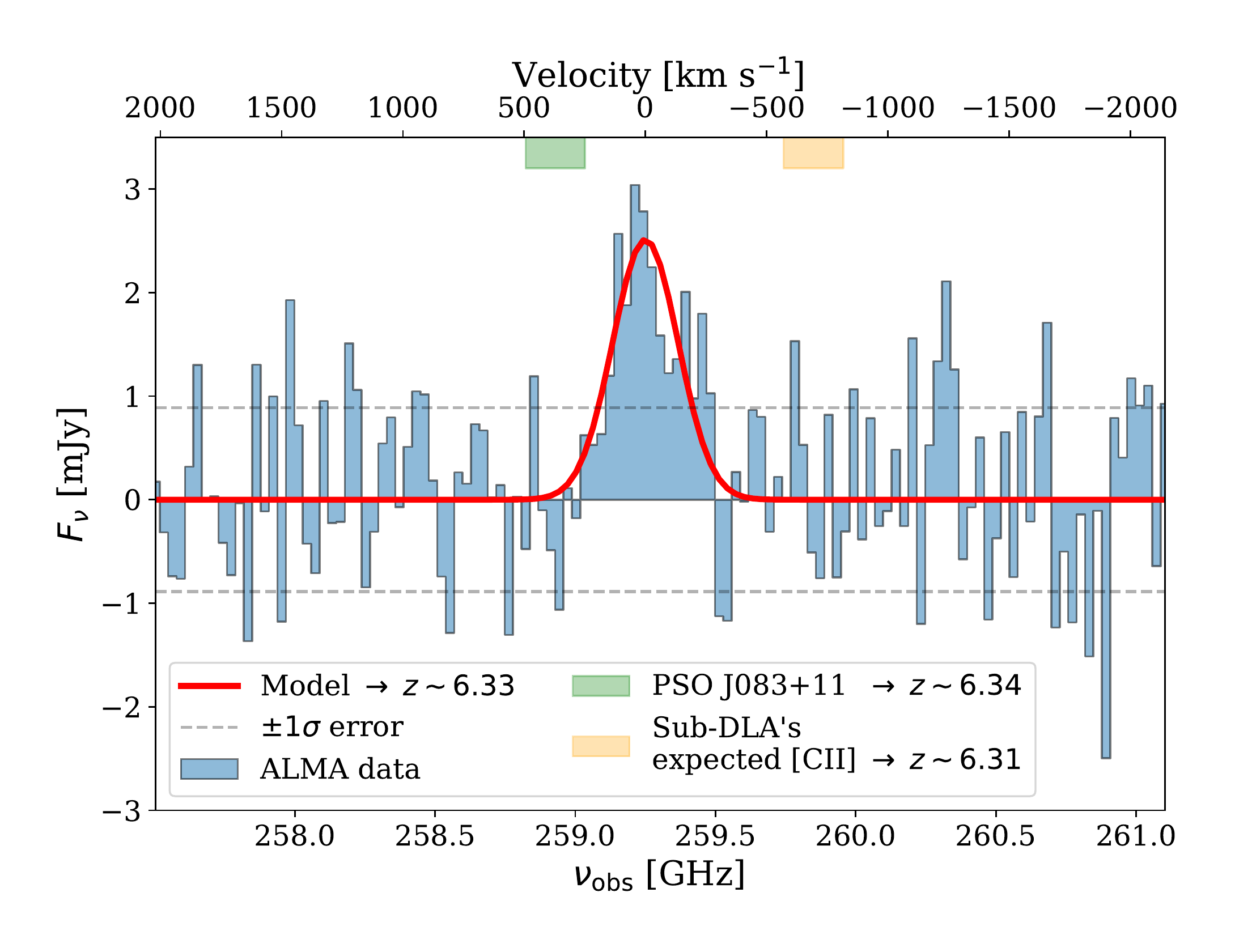}
	\caption{
		\textit{Left panel}: Velocity-integrated [\cii] map of the sky region around \PSOJ.
		We find a companion galaxy in the southwest direction from the central quasar, with a projected angular separation of 2\farcs88, or equivalent to $\sim 16$~kpc at $z=6.3309$.
		The ALMA synthesized beam's size is presented at the bottom left. 
		The solid black lines denote the [\cii] velocity-integrated flux contours with [3, 4, 5, 7, 9, 12, 15]~$\times \sigma$, where $\sigma = 0.053$~Jy~beam$^{-1}$~km~s$^{-1}$.
		Contours with negative values are shown with dashed black lines.		
		For extracting the spectrum and calculating the total [\cii] flux density of the companion galaxy, we use a circular aperture (see dotted circles) with a radius of 1\arcsec\ (equivalent to 5.5~kpc).
		\textit{Right panel}: The companion galaxy's [\cii] emission at the observed frequency of $\nu_\mathrm{obs} = 259.25$~GHz.
		The observed flux density, its associated $\pm 1\sigma$ uncertainty, and the fitted Gaussian model are shown with the blue, dashed gray, and solid red lines, respectively.
		The velocity axis centered at $z = 6.3309$ is presented at the top.
		The expected central wavelengths of [\cii] lines from the sub-DLA at $z=6.314$ and \PSOJ\ at $z=6.3401$ discussed in the main text are marked with the orange and green areas, respectively.
	}
	\label{fig:cii_mom0}
\end{figure*}

Due to the intrinsic faintness of $z\gtrsim2$ DLAs, finding the associated emission from their host galaxies is a challenging task \citep[e.g.,][]{2006ApJ...636...30K,2015MNRAS.446.3178F}.
To date, there are a few tens of cases where those DLA host galaxies are successfully detected, either at optical/near-infrared or far-infrared wavelengths \citep[e.g.,][]{2017MNRAS.469.2959K,2018MNRAS.479.2126F,2018ApJ...856L..23K,2018MNRAS.474.4039M,2019MNRAS.482L..65K,2019ApJ...870L..19N,2020Natur.581..269N}.
In this section, we attempt to search for the galaxy's emission associated with the sub-DLA toward \PSOJ\ using ALMA (C43-4 array configuration) data taken by \cite{2020ApJ...903...34A}.
At $z\sim 6.3$, the [\cii]~158~$\mu$m atomic fine-structure line is redshifted into the wavelengths covered by ALMA band-6, which is beneficial for identifying companion galaxy near \PSOJ.
We refer the reader to see \citet{2020ApJ...903...34A} for the details on the data reduction procedure.
The final reduced data cube covers 257.5--261.1~GHz spectral window and has a synthesized beam of 0\farcs42~$\times$~0\farcs37, 30~MHz channel width, and rms noise level of $\sim0.24$ mJy~beam$^{-1}$.

\begin{figure*}[htb!]
	\centering
	\epsscale{1.17}
	\plottwo{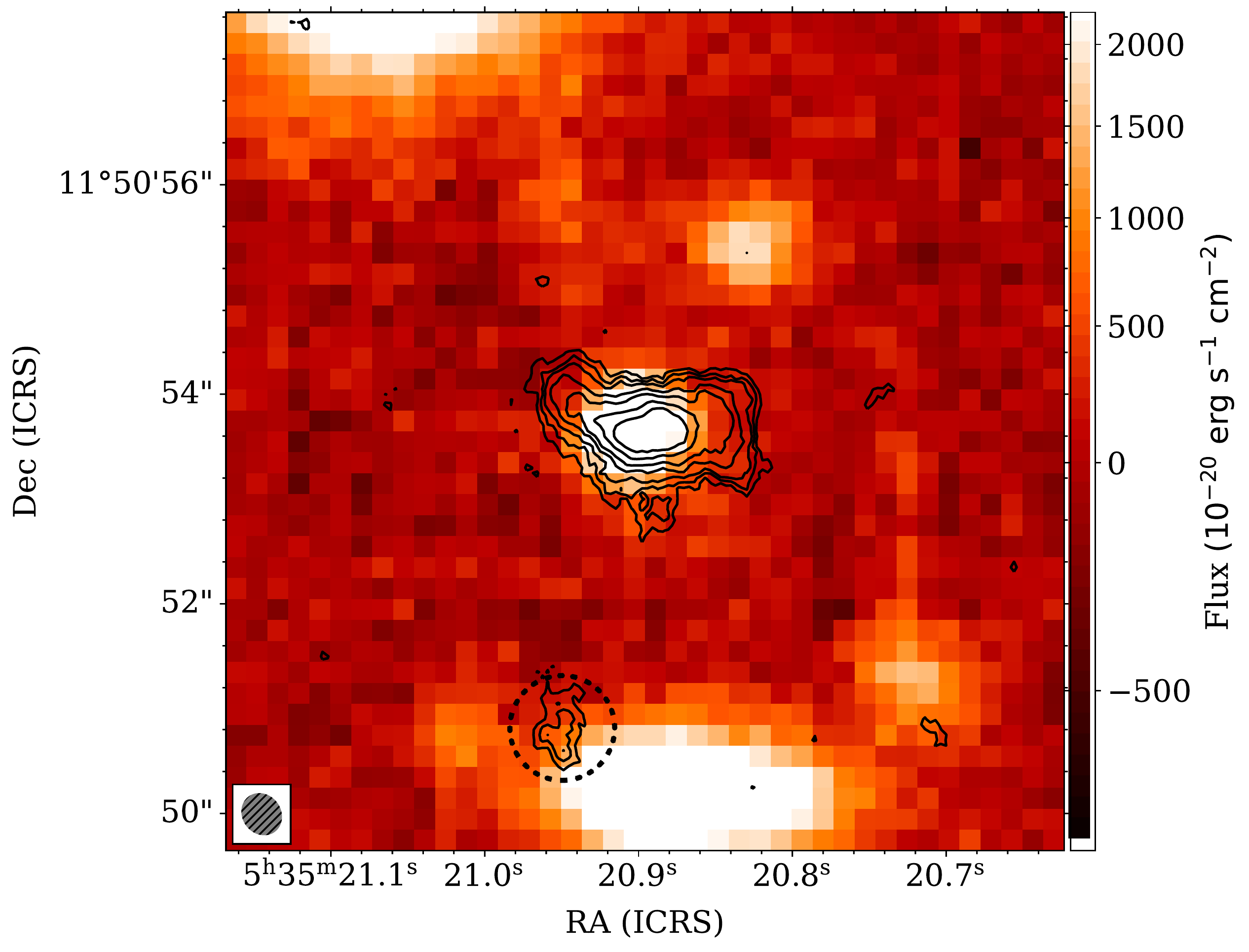}{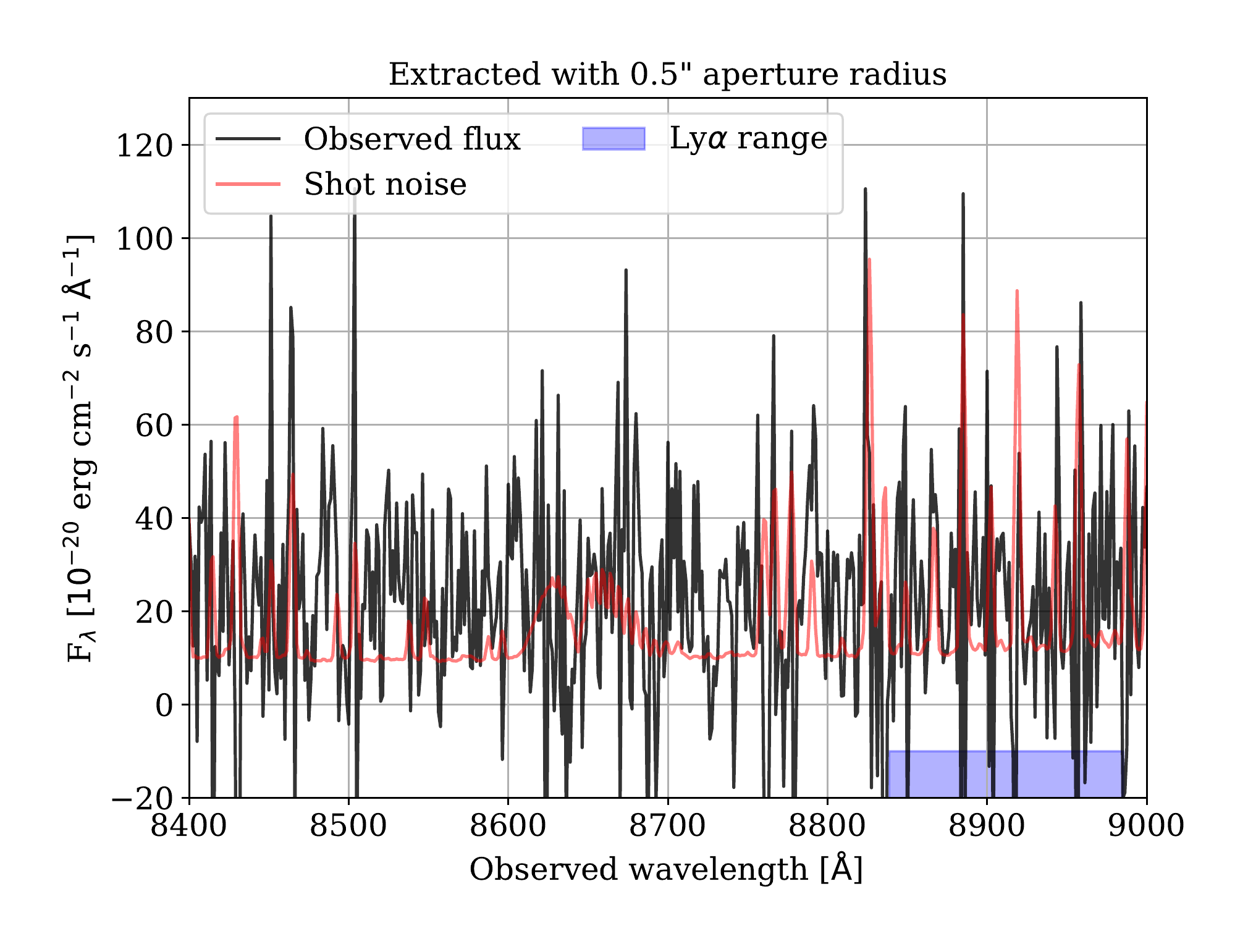}
	\caption{
		\textit{Left panel}: MUSE image centered around \PSOJ, created by collapsing the $\lambda_\mathrm{obs}=$ 8000--9000~\AA\ layers.
		An overlay of [\cii] emission from ALMA data is shown with a black contour.
		We then extract the spectrum at the region of interest using an aperture of 0\farcs5 radius (dotted circles).
		\textit{Right panel}: MUSE spectrum of J083.8372+11.8474 (black line) -- i.e., a $z=6.3309$ companion galaxy near \PSOJ\ --  and its associated shot noise (red line).
		We do not find any prominent emission lines in the wavelength region where the \Lya\ emission from the companion galaxy is expected (blue area).
		A faint continuum emission is seen, which originates from the adjacent bright galaxy at $z\approx0.29$ in the southeast direction.
	}
	\label{fig:comp_specmuse}
\end{figure*}

Through a visual inspection of the spectral channels, we detected emission of a companion galaxy, J083.8372+11.8474, in the southwest direction from \PSOJ, with a projected angular distance of 2\farcs88, or equivalent to a projected physical separation of $\sim 16$~kpc.
To visualize it, we created a moment-zero map\footnote{For details on the equation used for computing moment map, see \url{https://spectral-cube.readthedocs.io/en/latest/moments.html}.} of [\cii]~158~$\mu$m with a velocity width of 800~km~s$^{-1}$ centered at $\nu_\mathrm{obs} = 259.25$ GHz.
The result is shown in the left panel of Figure~\ref{fig:cii_mom0}, where we can see that the companion galaxy is observed as a marginally resolved single blob with a size of 0\farcs89~$\times$~0\farcs37, or corresponds to a physical extent of 4.9~kpc~$\times$~2.1~kpc at its redshift.
The size above is equivalent to the major and minor axis FWHMs, calculated by fitting a 2D Gaussian function to the velocity-integrated [\cii] map.
After that, we extracted the spectrum around the companion galaxy location using a circular aperture with a radius of 1\arcsec.
This particular aperture size was selected because there is no apparent emission beyond this region.
Aforementioned analysis led us to the discovery of a [\cii]\,158~$\mu$m emission at $z=6.3309\pm0.0004$ with the observed frequency of $\nu_\mathrm{obs} = 259.25\pm0.02$~GHz.
Note that the underlying continuum emission is not detected, where the estimated $3\sigma$ upper limit of flux density is 27.99~$\mu$Jy for 1\arcsec\ aperture radius.
By modeling the line with a 1D Gaussian function, we obtained an integrated [\cii] flux of $0.61\pm0.01$~Jy~km~s$^{-1}$ and FWHM of $318\pm58$~km~s$^{-1}$.

We subsequently calculate the [\cii] line luminosity following \cite{2013ARA&A..51..105C} prescription (see also Equations 15--16 in \citealt{2020ApJ...903...34A}), where the resulting value is $L_\mathrm{[CII]} = 6.22\pm0.14 \times 10^{8} L_\odot$.
After that, the star formation rate (SFR) can be estimated employing the known SFR--$L_\mathrm{[CII]}$ scaling relation for the high-$z$ galaxies \citep{2014A&A...568A..62D}, where we obtain SFR~$=71~M_\odot$~yr$^{-1}$.
It is also important to note that the relation above contains a systematic uncertainty of a factor of $\sim2.5$, which consequently makes the derived SFR has a range of $29$--179~$M_\odot$~yr$^{-1}$.

Previously, \cite{2019ApJ...870L..19N} reported [\cii] emission from a sample of four DLAs at $z\sim4$.
Those DLA host galaxies have relatively large projected separation from the central quasar (16--45~kpc), luminosities of $L_\mathrm{[CII]} = 0.36$--30~$\times 10^{8} L_\odot$, and SFRs of 7--110~$M_\odot$~yr$^{-1}$.
The companion galaxy we found here has similar [\cii] line properties as the aforementioned DLAs.
However, the [\cii]-based redshift of \PSOJ's companion galaxy significantly differs from the sub-DLA's redshift ($z_\mathrm{DLA}=6.314$) estimated via the centroids of rest-frame ultraviolet metal absorption lines.
The calculated velocity offset between those two sources is $\Delta V=692$~km~s$^{-1}$, or about two times the FWHM of [\cii], which make them more likely to be two unrelated galaxies. So, we do not identify any emission by the sub-DLA host galaxy.

We were also not able to confidently identify the optical counterpart of J083.8372+11.8474.
We searched by inspecting the MUSE image centered around \PSOJ\ with an overlay of [\cii] emission from ALMA data (see Figure~\ref{fig:comp_specmuse}).
We also extract the corresponding optical spectrum using an aperture with a 0\farcs5 radius.
There are no prominent emission lines detected around $\lambda_\mathrm{obs}=$ 8839--8985~\AA, where the \Lya\ emission from \PSOJ's companion galaxy is expected.
The apparent faint continuum emission in the spectrum most likely originates from the adjacent bright galaxy at $z\approx0.29$.

\clearpage

\bibliography{biblio}{}
\bibliographystyle{aasjournal}



\end{document}